\documentclass[11pt]{article}

\usepackage{amsfonts,amsmath,amssymb,bm,graphicx,hyperref,a4wide,cite,mathabx,subfigure}

\begin{document}

\title{\textbf{Spin oscillations of neutrinos scattered off a rotating black hole}}

\author{Maxim Dvornikov\thanks{maxdvo@izmiran.ru} 
\\
\small{\ Pushkov Institute of Terrestrial Magnetism, Ionosphere} \\
\small{and Radiowave Propagation (IZMIRAN),} \\
\small{108840 Troitsk, Moscow, Russia}}

\date{}

\maketitle

\begin{abstract}
Spin oscillations of neutrinos, gravitationally scattered
off a black hole (BH), are studied. The cases of nonrotating and rotating BHs are analyzed. We derive the analytic expressions for the transition and survival
probabilities of spin oscillations when neutrinos interact
with these gravitational backgrounds. 
The obtained transition probabilities depend on the impact parameter, as well as the neutrino energy and the particle mass. We find that there is a possibility of spin oscillations of ultrarelativistic neutrinos scattering off a rotating BH.
Then, considering the neutrino scattering off BH surrounded
by background matter, we derive the effective Schr\"{o}dinger equation for spin oscillations. 
The numerical solution of this equation is obtained in the case
of a supermassive BH with a realistic accretion disk. Spin effects turn out to be negligible in the neutrino scattering in the Schwarzschild metric. In the Kerr metric, we find that
the observed neutrino fluxes can be reduced almost 10\% because of
spin oscillations when ultrarelativistic neutrinos experience gravitational scattering.
The neutrino interaction with an accretion disk results in the additional
modification of the intensities of outgoing neutrino fluxes. We consider the applications of the obtained results for the neutrino astronomy.
\end{abstract}

\maketitle

\section{Introduction}

The recent successful studies of oscillations of accelerator~\cite{Aga18} and atmospheric~\cite{Abe18} neutrinos demonstrate that the masses of these particles are nonzero and there is a mixing between different neutrino flavors. This fact is the direct indication to the physics beyond the standard model. One can study transitions between neutrinos belonging to different flavors, or neutrino flavor oscillations. This type of oscillations is the most plausible solution of the solar neutrino problem~\cite{Gir18}.

However, in the present work, we shall concentrate on neutrino spin oscillations, which were first proposed in Ref.~\cite{FujShr80}. Neutrinos are left polarized particles in the standard model. If such a neutrino changes its polarization under the action of some external backgrounds, it cannot be observed since right neutrinos are sterile particles. This process leads to the suppresion of the emitted flux of left neutrinos. The study of neutrino spin and spin-flavor oscillations in various external fields can provide a valuable information about neutrino magnetic moments~\cite{GiuStu15}. 

The neutrino interactions with external fields, e.g., the electroweak interaction 
with background matter~\cite{Bil18}, are known to influence the process of neutrino
oscillations. The neutrino interaction with gravitational fields of astrophysical objects, despite of its weakness, can also change the dynamics of neutrino oscillations. Earlier, we examined quasiclassically neutrino spin oscillations under the influence of external fields in curved spacetime in Refs.~\cite{Dvo06,Dvo13,Dvo19}, where both static metrics and time dependent backgrounds,
like a gravitational wave, were studied. The quantum description of the fermion
spin evolution in curved spacetime was developed in Ref.~\cite{ObuSilTer17}.

We should make a remark on the studies of neutrino spin oscillations in gravitational fields in Refs.~\cite{Dvo06,Dvo13,SorZil07,AlaNod15}. In these works, neutrinos were supposed to be gravitationally captured by a massive object, like a black hole (BH).  In this case, the effect of spin oscillations is not observable since a neutrino detector is typically outside the BH region. Thus 
one has to consider neutrino spin oscillations in the neutrino gravitational scattering, when one can potentially measure the helicities of both incoming and outgoing particles. Note that flavor oscillations of neutrinos gravitationally scattered off BH were studied in Refs.~\cite{CroGiuMor04,AleClo18}.
 
Before the recent observation of the event horizon silhouette 
of a supermassive BH (SMBH)~\cite{Aki19}, which is the main motivation for the present work, all the experimental manifestations of the general relativity, including the direct  detection of gravitational waves~\cite{Abb19}, corresponded to the weak field limit. The bright halo around this silhouette is formed by photons emitted by the accretion disk around SMBH. The size of the event horizon silhouette, computed in Ref.~\cite{DokNazSmi19}, turns out to be in the agreement with the  prediction of the general relativity.

However, besides photons, a significant
flux of neutrinos is expected to be emitted by dense and hot matter of  an accretion disk. These neutrinos were found in Ref.~\cite{WanJan12} to modify the $r$-process nucleosynthesis in the vicinity of BH surrounded by an accretion disk. The spin of emitted neutrinos can precess in the gravitational field of a central BH.
In this work, we shall examine how a strong gravitational field of
BH and the neutrino interaction with an accretion disk can cause their helicity change.

The neutrino gravitational scattering was studied recently~\cite{Cor15}, in order to compute the size and the shape of the BH shadow formed
by these particles~\cite{StuSch19}. In our work, we shall study neutrino spin oscillations in the gravitational scattering. This process is expected to suppress the flux of particles measured
with a terrestrial neutrino telescope.
 
It was found in Ref.~\cite{CunHer18} that photons, which form a bright halo around the event horizon silhouette of BH, interact with both its gravitational field and plasma which surrounds BH. This interaction
with plasma leads to the modification of the size and the shape of the BH shadow. In the present work, the role of the neutrino interaction with background matter, e.g., with
an accretion disk, for the detected flux of gravitationally
scattered neutrinos is examined.


In this our work, we continue our studies of neutrino spin oscillations
in Refs.~\cite{Dvo06,Dvo13,Dvo19}. We start in Sec.~\ref{sec:GRAV}
with the analysis of the neutrino spin evolution when a particle gravitationally
scatters off BH. We find the general expressions
for the transition and survival probabilities. Then, we apply our results for the description of spin oscillations in the neutrino scattering off a nonrotating BH in Sec.~\ref{sec:SCHWARZ} and a rotating one in Sec.~\ref{sec:KERR}. In Sec.~\ref{sec:MATT}, we formulate the effective Schr\"{o}dinger equation
for neutrino spin oscillations in the scattering off BH surrounded
by background matter. We study astrophysical applications in Sec.~\ref{sec:APPL}.
In particular, we consider the effect of spin oscillations on the
measured neutrino fluxes when particles scatter off SMBH with a realistic
accretion disk. The situations of nonrotating and rotating BHs are studied. Finally, in Sec.~\ref{sec:DISC}, we discuss our
results. We remind how a scalar particle
moves in the Schwarzschild metric in Appendix~\ref{sec:PARTM} and in Kerr metric in Appendix~\ref{sec:PARTMKERR}.

\section{Neutrino spin evolution in gravitational scattering\label{sec:GRAV}}

In this section, we study how the spin of a neutrino evolves when
a particle scatters off BH. First, we derive the general expressions for the transition and survival probabilities. Then, we apply this result for spin oscillations in the Schwarzschild and Kerr metrics. In case of the Kerr metric, we study the neutrino motion in the equatorial plane only. In these situations, one can analytically solve the spin evolution equation and obtain the probabilities for spin oscillations in quadratures.  

In Refs.~\cite{Dvo06,Dvo13}, we found that the neutrino invariant
spin $\bm{\zeta}$, defined in a locally Minkowskian frame, evolves
in an external gravitational field as
\begin{equation}\label{eq:spinevgen}
  \frac{\mathrm{d}\bm{\zeta}}{\mathrm{d}t}=2(\bm{\zeta}\times\bm{\Omega}_{g}),
\end{equation}
where $t$ is the time in world coordinates and $\bm{\Omega}_{g}$ is the vector accounting for the gravity contribution. If a neutrino interacts
with a Schwarzschild or Kerr BH, $\bm{\Omega}_{g}$ in Eq.~(\ref{eq:spinevgen})
has only one nonzero component~\cite{Dvo06,Dvo13}, $\bm{\Omega}_{g}=(0,\Omega_{2},0)$.

We are interested in neutrino spin oscillations, i.e. in the change
of the neutrino helicity, $h=(\bm{\zeta}\mathbf{u})/|\mathbf{u}|$,
where $\mathbf{u}$ is the spatial part of the neutrino four velocity
in the locally Minkowskian frame. Therefore, besides the study of
the neutrino spin in Eq.~(\ref{eq:spinevgen}), we should account
for the evolution of $\mathbf{u}$.

In principle, we can avoid simultaneous tracking of both $\bm{\zeta}$ and $\mathbf{u}$. However, we should fix the initial and final neutrino polarizations. As we shall see in Secs.~\ref{sec:SCHWARZ} and~\ref{sec:KERR}, at $r\to\infty$, $\mathbf{u}(t\to\pm\infty)=\mathbf{u}_{\pm\infty}=\left(\pm\left[E^{2}-m^{2}\right]^{1/2}/m,0,0\right)$, where $r$ is the distance between the BH center and a neutrino, $E$ is the neutrino energy, which is the integral of motion in the considered metrics, and $m$ is the neutrino mass. Thus the asymptotic neutrino motion happens along the first axis in the locally Minkowskian frame. In this frame, an incoming neutrino ($t\to-\infty$)
propagates oppositely the first axis. An outgoing particle ($t\to+\infty$) moves along
this axis.

Since only $\Omega_{2}\neq0$, the nonzero neutrino spin components
are $\zeta_{1,3}\neq0$, and $\zeta_{2}=0$. It is convenient to represent
\begin{equation}\label{eq:zeta13alpha}
  \zeta_{1}=\cos\alpha,\quad\zeta_{3}=\sin\alpha,
\end{equation}
where $\alpha$ is the angle between $\bm{\zeta}$ and the positive direction of the first axis in the locally Minkowskian frame. Now we have to specify the initial condition for Eq.~(\ref{eq:spinevgen}).
We suppose that, initially, at $r\to\infty$, an incoming
neutrino is left polarized, i.e. the helicity is negative, $h_{-\infty}=(\bm{\zeta}_{-\infty}\mathbf{u}_{-\infty})/|\mathbf{u}_{-\infty}|=-1$.
Accounting for the expression for $\mathbf{u}_{-\infty}$ above, we
get that $\zeta_{-\infty1}=1$ and $\zeta_{-\infty3}=0$, or $\alpha_{-\infty}=0$
in Eq.~(\ref{eq:zeta13alpha}).

The helicity of an outgoing neutrino has the form, $h_{+\infty}=(\bm{\zeta}_{+\infty}\mathbf{u}_{+\infty})/|\mathbf{u}_{+\infty}|$,
where $\bm{\zeta}_{+\infty}=(\cos\alpha_{+\infty},0,\sin\alpha_{+\infty})$
and $\mathbf{u}_{+\infty}$ is given above. Using Eq.~(\ref{eq:zeta13alpha}),
we get that $h_{+\infty}=\cos\alpha_{+\infty}$. The transition $P_{\mathrm{LR}}$
and survival $P_{\mathrm{LL}}$ probabilities for neutrino spin oscillations
are
\begin{equation}\label{eq:Pgen}
  P_\mathrm{LR,LL}=\frac{1}{2}(1\pm \cos\alpha_{+\infty}),
\end{equation}
where the upper sign stays for $P_{\mathrm{LR}}$ and the lower one
for $P_{\mathrm{LL}}$.

\subsection{Schwarzschild metric\label{sec:SCHWARZ}}

First, we study the neutrino motion in the field of a nonrotating BH.
Using the spherical coordinates $(r,\theta,\phi)$, the interval in
this case has the form~\cite[p.~284]{LanLif71},
\begin{equation}\label{eq:intschw}
  \mathrm{d}\tau^{2}=
  \left(
    1-\frac{r_{g}}{r}
  \right)
  \mathrm{d}t^{2}-
  \left(
    1-\frac{r_{g}}{r}
  \right)^{-1}
  \mathrm{d}r^{2}-r^{2}(\mathrm{d}\theta^{2}+\sin^{2}\theta\mathrm{d}\phi^{2}),
\end{equation}
where $r_{g}=2M$ is the gravitational radius,
and $M$ is the BH mass. Since the Schwarzschild metric in Eq.~(\ref{eq:intschw})
is spherically symmetric, we can take that a neutrino moves in the
equatorial plane with $\theta=\pi/2$, i.e. $\mathrm{d}\theta=0$.

The nonzero component of $\bm{\Omega}_{g}$ in the Schwarzschild metric has the form~\cite{Dvo06},
\begin{equation}\label{eq:Omega2}
  \Omega_{2}=\frac{L}{2Er^{2}}
  \left(
    1-\frac{r_{g}}{r}
  \right)
  \left(
    -\sqrt{1-\frac{r_{g}}{r}}+
    \frac{U^{t}}{1+U^{t} \sqrt{1-r_g/r}}\frac{r_{g}}{2r}
  \right),
\end{equation}
where $L$ is the conserved
angular momentum of a neutrino. In Eq.~(\ref{eq:Omega2}), $U^{t} = \mathrm{d}t/\mathrm{d}\tau = E(1-r_g/r)^{-1}/m$ is the component of the four velocity $U^\mu$ in world coordinates.

The expression for $\mathbf{u}$ in the Schwarzschild metric was also obtained in Ref.~\cite{Dvo06},
\begin{equation}\label{eq:uexpl}
  \mathbf{u}= 
  \left(
    \pm\frac{1}{m}
    \left[
      E^{2}-m^{2}
      \left(
        1-\frac{r_{g}}{r}
      \right)
      \left(
        1+\frac{L^{2}}{m^{2}r^{2}}
      \right)
    \right]^{1/2},
    0,\frac{L}{mr}
  \right),
\end{equation}
where the signs $\pm$ stay for outgoing and incoming neutrinos respectively
[see Eq.~(\ref{eq:eqmtr})]. Using Eq.~\eqref{eq:uexpl} at $r\to\infty$, we obtain the expression for $\mathbf{u}_{\pm\infty}$ proposed above.


Then we use Eqs.~(\ref{eq:spinevgen}), (\ref{eq:Omega2}), and~(\ref{eq:eqmtr}) to determine the evolution of $\alpha$ in Eq.~(\ref{eq:zeta13alpha}). It obeys the equation,
\begin{equation}\label{eq:F}
  \frac{\mathrm{d}\alpha}{\mathrm{d}r}= 
  \pm\frac{L}{mr^{2}}
  \frac{\frac{E}{m}
    \left(\tfrac{3r_{g}}{2r}-1
  \right)-
  \left(
    1-\tfrac{r_{g}}{r}
  \right)^{3/2}}{
  \tfrac{E}{m}+
    \left(
      1-\tfrac{r_{g}}{r}
    \right)^{1/2}}
  \left[
    \frac{E^{2}}{m^{2}}-
      \left(
        1-\frac{r_{g}}{r}
      \right)
    \left(
      1+\frac{L^{2}}{m^{2}r^{2}}
    \right)
  \right]^{-1/2},
\end{equation}
where the signs $\pm$ stay for outgoing and incoming neutrinos. Now,
accounting for the initial condition $\alpha_{-\infty}=0$
and the fact that $\alpha_{+\infty}$ is twice the angle corresponding
to the minimal distance between a neutrino and BH, we get that $\alpha_{+\infty}$ reads,
\begin{equation}\label{eq:aobslim}
  \alpha_{+\infty}=\int_{x_{m}}^{\infty} F_\mathrm{S}(x)\mathrm{d}x,
  \quad
  F_\mathrm{S}(x) = \frac{y}{x\sqrt{(x-1)R_\mathrm{S}(x)}}
  \frac{(3-2x)\sqrt{x}-2\gamma^{-1}(x-1)^{3/2}}{\sqrt{x}+\gamma^{-1}\sqrt{x-1}},
\end{equation}
where $y=b/r_{g}$, $b=L/E\sqrt{1-\gamma^{-2}}$ is the impact parameter, $\gamma = E/m$ in the Lorentz factor at the infinity, and $x_{m}$
is the maximal root of the equation
\begin{equation}\label{eq:eqtosolve}
  R_\mathrm{S}(x) = x^{3}+\frac{\gamma^{-2}}{1-\gamma^{-2}}x^2 - y^{2}(x-1)=0.
\end{equation}
Note that $y > y_0$, where
\begin{equation}\label{eq:ycrit}
  y_0 = \frac{1}{2\sqrt{2}(1-\gamma^{-2})}
  \left[
    9(3 + \sqrt{9 - 8\gamma^{-2}}) - 4\gamma^{-2}(9+2\sqrt{9 - 8\gamma^{-2}}) +
    8 \gamma^{-4}
  \right]^{1/2}.
\end{equation}
for a neutrino not to fall to BH (see Appendix~\ref{sec:PARTM}). For ultrarelativistic neutrinos with $\gamma\gg 1$, we obtain the well known result, $y_0 = 3\sqrt{3}/2$.

First, let us analyze $\alpha_{+\infty}$ in the limit $\gamma\gg 1$. Equation~\eqref{eq:aobslim} takes the form,
\begin{equation}\label{eq:abiggamma}
  \alpha_{+\infty}=y\int_{x_{m}}^{\infty}
  \frac{\mathrm{d}x(3-2x)}{x\sqrt{(x-1)R_\mathrm{S}(x)}},
\end{equation}
where $R_\mathrm{S}(x)\to x^{3}-y^{2}(x-1)$. 
%
Basing on 
Eq.~\eqref{eq:abiggamma}, one can argue that $\alpha_{+\infty} = -\pi$ for any $3\sqrt{3}/2 \leq y < \infty$. Hence, using Eq.~\eqref{eq:Pgen}, we get that $P_\mathrm{LR} = (1+\cos\alpha_{+\infty})/2 = 0$ at $E\gg m$ for the arrbirtary impact parameter. It means that there is no spin flip of ultrarelativistic neutrinos when they scatter off a Schwarzschild BH. This fact is in agreement with the results of Ref.~\cite{DolDorLas06}.

Now we can correct the result of Ref.~\cite{Dvo20}, where the nonzero transition probability for spin oscillations of ultrarelativistic neutrinos scattering off a  Schwarzschild BH was obtained. That incorrect result is a consequence of the extra factor $1/U^t$ in Eq.~\eqref{eq:spinevgen} used in Ref.~\cite{Dvo20}. The vector $\bm{\Omega}_g$ in Eq.~\eqref{eq:spinevgen} already accounts for the change of the proper time $\tau$ to the world time $t$.

At the end of this section, we return to the case of the finite neutrino energy. Considering the situation $y=y_0$, the roots of Eq.~\eqref{eq:eqtosolve} have the form,
\begin{align}\label{eq:ragamma}
  x_0 = & x_1 = \frac{3 + \sqrt{9 - 8\gamma^{-2}}-4\gamma^{-2}}{4(1-\gamma^{-2})},
  \notag
  \\
  x_2 = & - \frac{9(3 + \sqrt{9 - 8\gamma^{-2}}) - \gamma^{-2} (27 + 
   5\sqrt{9 - 8\gamma^{-2}}) + 4\gamma^{-4}}
  {(1-\gamma^{-2})(9 + 3\sqrt{9 - 8\gamma^{-2}} - 4\gamma^{-2})}.
\end{align}
When these $x_{0,1,2}$ in Eq.~\eqref{eq:ragamma} are used in Eq.~\eqref{eq:aobslim}, the value of $\alpha_{+\infty}$ turns out to be infinite. Thus one cannot express $\alpha_{+\infty}$ as a power series of $\gamma^{-1}$ at $y=y_0$. It means that the neutrino spin makes infinite revolutions with respect to the neutrino velocity when a particle asymptotically approaches a nonrotating BH.

\begin{figure}
  \centering
  \subfigure[]
  {\label{1a}
  \includegraphics[scale=.4]{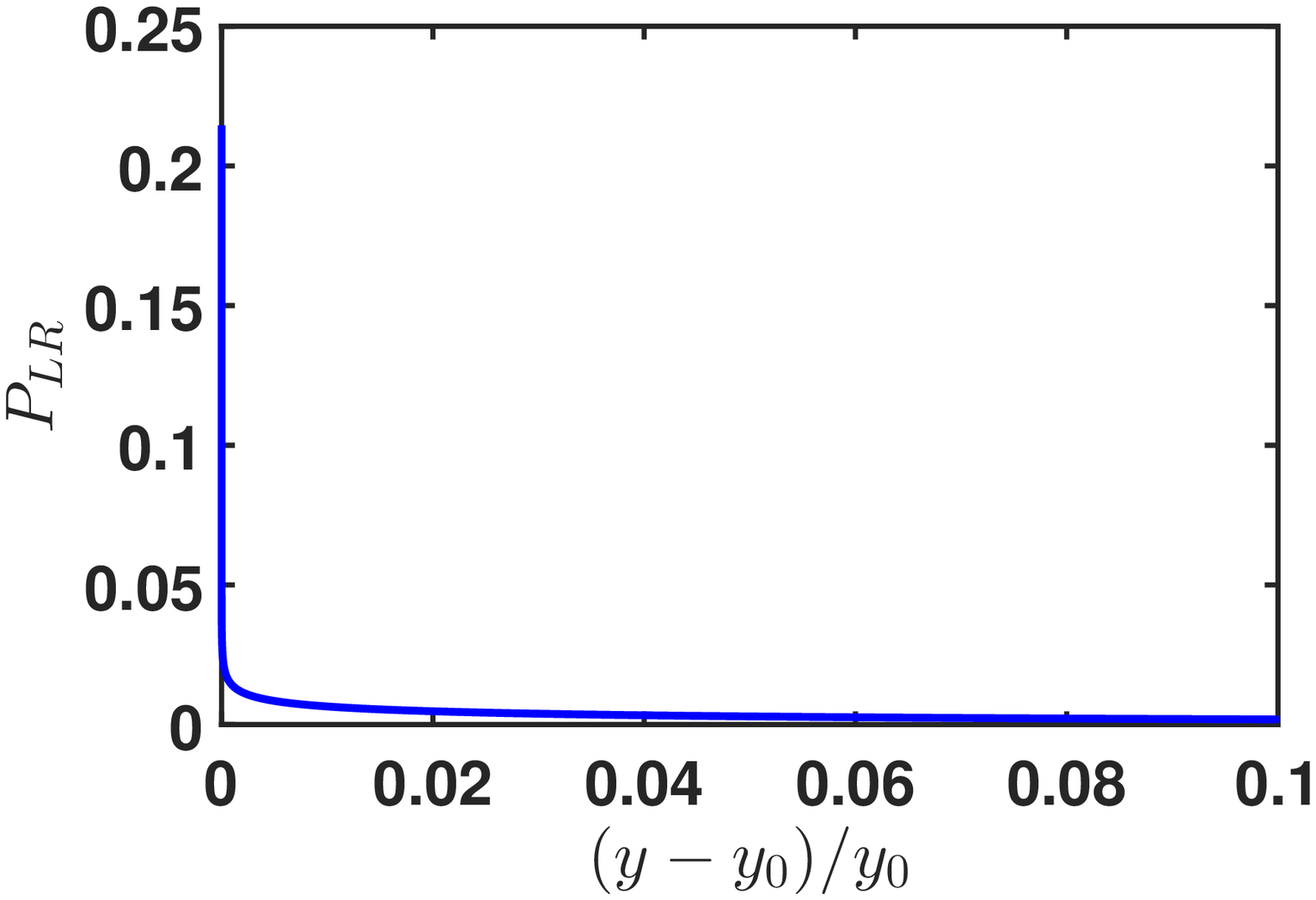}}
  \hskip-.6cm
  \subfigure[]
  {\label{1b}
  \includegraphics[scale=.4]{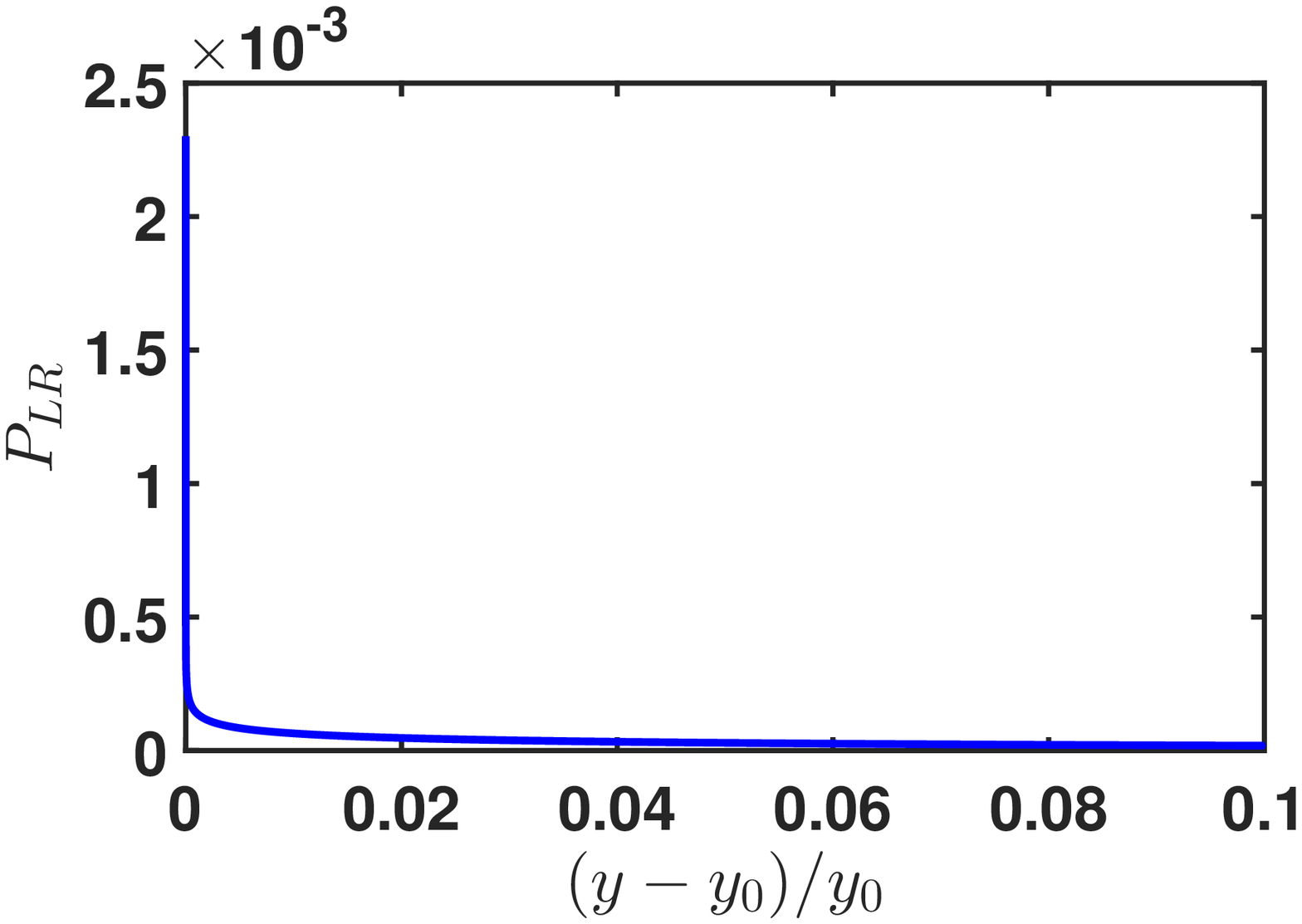}}
  \protect
  \caption{The transition probability $P_\mathrm{LR}$
  of spin oscillations versus the dimensioless impact parameter $y$
  for neutrinos scattering off a Schwarzschild BH for different neutrino energies.
  (a)~$E = 10 m$, (b)~$E = 100 m$.\label{fig:PLRSchwarz}}
\end{figure}

In Fig.~\ref{fig:PLRSchwarz}, we show the transition probabilities of spin oscillations of massive neutrinos scattered off a nonrotating BH versus the impact parameter. The maximal transition probability is reached when $y\to y_0$. One can see that the greater $\gamma = E/m$ is the smaller $P_\mathrm{LR}$ is. This behavior of $P_\mathrm{LR}$ confirms our result that there are no spin oscillations in scattering of massless neutrinos in the Schwarzschild metric.

%

\subsection{Kerr metric\label{sec:KERR}}

Now we study spin evolution of neutrinos scattered off a rotating BH. The space-time in this case is described by the Kerr metric. In Boyer-Lindquist coordinates $(t,r,\theta,\phi)$, this metric has the form~\cite{Rez16},
\begin{equation}\label{eq:Kerrmetr}
  \mathrm{d}\tau^{2}=
  \left(
    1-\frac{rr_{g}}{\Sigma}
  \right)
  \mathrm{d}t^{2}+
  2\frac{rr_{g}a\sin^{2}\theta}{\Sigma}\mathrm{d}t\mathrm{d}\phi-
  \frac{\Sigma}{\Delta}\mathrm{d}r^{2}-
  \Sigma\mathrm{d}\theta^{2}-
  \frac{\Xi}{\Sigma}\sin^{2}\theta\mathrm{d}\phi^{2},
\end{equation}
where
\begin{equation}\label{eq:Kerrmetrparam}
  \Delta=r(r-r_{g})+a^{2},
  \quad
  \Sigma=r^{2}+a^{2}\cos^{2}\theta,
  \quad
  \Xi=
  \left(
    r^{2}+a^{2}
  \right)
  \Sigma+rr_{g}a^{2}\sin^{2}\theta.
\end{equation}
The parameter $a$ in Eqs.~\eqref{eq:Kerrmetr} and~\eqref{eq:Kerrmetrparam} can be in the range, $0<a<r_g/2$. The angular momentum of BH is $J=Ma$.

We study a neutrino moving in the equatorial plane of a Kerr BH with $\theta=\pi/2$ and $\mathrm{d}\theta=0$. As in the Schwarzschild case, we also have two integrals of motion: the energy $E$ and the neutrino angular momentum $L$. However, we should consider two cases $L>0$ and $L<0$ in the Kerr metric.

The vector $\bm{\Omega}_g$ has the nonzero component~\cite{Dvo13},
\begin{equation}\label{eq:Omega2Kerr}
  \Omega_{2}=\frac{1}{2U^{t}}
  \left(
    b_{2}+\frac{e_{3}u^{1}-e_{1}u^{3}}{1+u^{0}}
  \right).
\end{equation}
where
\begin{align}\label{eq:b2e13}
  b_{2} = & -\frac{\sqrt{r(r-r_{g})+a^{2}}}{2r^{2}\sqrt{r[r^{3}+a^{2}(r+r_{g})]}}
  [(2r^{3}-r_{g}a^{2})U^{\phi}+r_{g}aU^{t}],
  \notag
  \\
  e_{1} = & \frac{r_{g}}{2r^{2}}
  \frac{[a(3r^{2}+a^{2})U^{\phi}-(r^{2}+a^{2})U^{t}]}{\sqrt{r[r^{3}+a^{2}(r+r_{g})]}},
  \notag
  \\
  e_{3} = & \frac{ar_{g}}{2r}
  \frac{(3r^{2}+a^{2})U^{r}}{[r^{3}+a^{2}(r+r_{g})]\sqrt{r(r-r_{g})+a^{2}}},
\end{align}
are the components the gravi-magnetic and gravi-electric fields,
\begin{align}\label{eq:u013}
  u^{0} = & U^{t}\frac{\sqrt{r[r(r-r_{g})+a^{2}]}}{\sqrt{r^{3}+a^{2}(r+r_{g})}},
  \notag
  \\
  u^{1} = & \frac{rU^{r}}{\sqrt{r(r-r_{g})+a^{2}}},
  \notag
  \\
  u^{3} = & \frac{(r^{3}+ra^{2}+r_{g}a^{2})U^{\phi}-ar_{g}U^{t}}
  {\sqrt{r[r^{3}+a^{2}(r+r_{g})]}},
\end{align}
are the components of the neutrino four velocity $u^a = (u^0,u^1,0,u^3)$ in the locally Minkowskian frame, and $U^\mu = (U^{t},U^r,0,U^\phi) = \mathrm{d}x^\mu/\mathrm{d}\tau$ is the neutrino four velocity in the world coordinates. For example,
\begin{equation}\label{eq:U0Kerr}
  U^{t}=\frac{[r^{3}+a^{2}(r+r_{g})]E-a L r_{g}}{mr[r(r-r_{g})+a^{2}]}.
\end{equation}
The expressions for $U^r$ and $U^\phi$ can be found on the basis of Eqs.~\eqref{eq:U0Kerr} and~\eqref{eq:eqmtrKerr}.

Equations~\eqref{eq:zeta13alpha} and~\eqref{eq:Pgen} are valid for the motion in the equatorial plane of a Kerr BH. Using Eqs.~\eqref{eq:b2e13}-\eqref{eq:U0Kerr} and applying the similar technique as in Sec.~\ref{sec:SCHWARZ}, we get the expression for the angle $\alpha_{+\infty}$, which determines the probabilities of spin oscillations, in the form,
\begin{align}
  \alpha_{+\infty}=&\int_{x_{m}}^{\infty} F_\mathrm{K}(x)\mathrm{d}x,
  \label{eq:alphaintKerr}
  \\
  F_\mathrm{K}(x) & =
  \frac{1}{x\sqrt{(1-\gamma^{-2})R_\mathrm{K}(x)[x(x-1)+z^2]}}
  \notag
  \\
  & \times
  \bigg\{
    - \frac{2x^2(x-1)y\sqrt{1-\gamma^{-2}} + 3zx^2 - z^2y\sqrt{1-\gamma^{-2}}+z^3}
    {\sqrt{x^3+z^2(x+1)-yz\sqrt{1-\gamma^{-2}}}}
    \notag
    \\
    & +
    \sqrt{1-\gamma^{-2}}[x^3+z^2(x+1)]^{-1/2}
    \notag
    \\
    & \times
    \Big(
      x^3+z^2(x+1)-yz\sqrt{1-\gamma^{-2}} 
      \notag
      \\
      & +
      \gamma^{-1}\sqrt{x[x(x-1)+z^2][x^3+z^2(x+1)]}
    \Big)^{-1}
    \notag
    \\
    & \times
    \Big[
      xy
      \big(
        x^4 - zx(3x-2)y\sqrt{1-\gamma^{-2}} - 2z^2x(1-x) - z^3y\sqrt{1-\gamma^{-2}}+z^4
      \big)
      \notag
      \\
      & +
      z(3x^2+z^2)\sqrt{1-\gamma^{-2}}R_\mathrm{K}(x)
    \Big]
  \bigg\},
  \label{eq:alphaKerr}
\end{align}
where $z = a/r_g$ and $x_m$ is the maximal root of the equation,
\begin{equation}\label{eq:RKerr}
  R_\mathrm{K}(x) = x^{3}+\frac{\gamma^{-2}}{1-\gamma^{-2}}x^2 + (z^{2}-y^2)x +
  \left(
    y - \frac{z}{\sqrt{1-\gamma^{-2}}}
  \right)^2 = 0.  
\end{equation}
One can check that, at $z \to 0$, $F_\mathrm{K}(x)\to F_\mathrm{S}(x)$ and $R_\mathrm{K}(x)\to R_\mathrm{S}(x)$, i.e., using Eqs.~\eqref{eq:alphaKerr} and~\eqref{eq:RKerr}, we reproduce the spin evolution in the Schwarzschild metric, studied in Sec.~\ref{sec:SCHWARZ}.

It should be noted that, while studying the scattering off a rotating BH, we should distinguish the cases of the direct scattering with $L>0$ and the retrograde one $L<0$ (see Fig.~\ref{fig:Kerrscatt} in Appendix~\ref{sec:PARTMKERR}). We can formally suggest that the impact parameter is positive in the retrograde scattering, but consider the expressions similar to Eqs.~\eqref{eq:alphaKerr} and~\eqref{eq:RKerr}, where all terms with odd powers of $y$ have opposite signs.

As in Sec.~\ref{sec:SCHWARZ}, it is interesting to study the case of ultrareletivistic particles. Using Eqs.~\eqref{eq:alphaKerr} and~\eqref{eq:RKerr} in the limit $\gamma\gg 1$, we get that
\begin{align}\label{eq:FKultrarel}
  F_\mathrm{K}(x) & \to
  \frac{1}{x\sqrt{R_\mathrm{K}(x)[x(x-1)+z^2][x^3+z^2(x+1)-yz]}}
  \notag
  \\
  & \times
  \Big\{
    - 2 x^2 (x-1) y - 3 z x^2 + z^2 y - z^3
    \notag
    \\
    & +
    [x^3+z^2(x+1)]^{-1/2}[x^3+z^2(x+1)-yz]^{-1/2}
    \notag
    \\
    & \times
    \big[
      x^5 y + z x^2 (3 x^3 + 5 y^2 - 6 x y^2) + 2 z^2 y x^2 (x  - 4) 
      \notag
      \\
      & +
      z^3 (4 x^3  - 2 x y^2  + 3 x^2 + y^2)
      + z^4 y (x - 2) + z^5(x + 1)
    \big]
  \Big\},
\end{align}
and
\begin{equation}\label{eq:RKultrarel}
  R_\mathrm{K}(x) \to x^{3} + (z^{2}-y^2)x + (y - z)^2.
\end{equation}

The simultaneous solution of the equations $R_\mathrm{K}(x)=0$ and $R_\mathrm{K}'(x)=0$ gives one the critical impact parameter,
\begin{equation}\label{eq:ycritKerr}
  y_{0}=4\cos^{3}
  \left[
    \frac{1}{3}\arccos(\mp2z)
  \right]\pm z,
\end{equation}
where the upper signs stay for the direct scattering and the lower ones for the retrograde scattering. Note that, we made that both $y_0$ in Eq.~\eqref{eq:ycritKerr} are positive. If $y<y_0$, a neutrino asymptotically falls to BH.

\begin{figure}
  \centering
  \subfigure[]
  {\label{2a}
  \includegraphics[scale=.4]{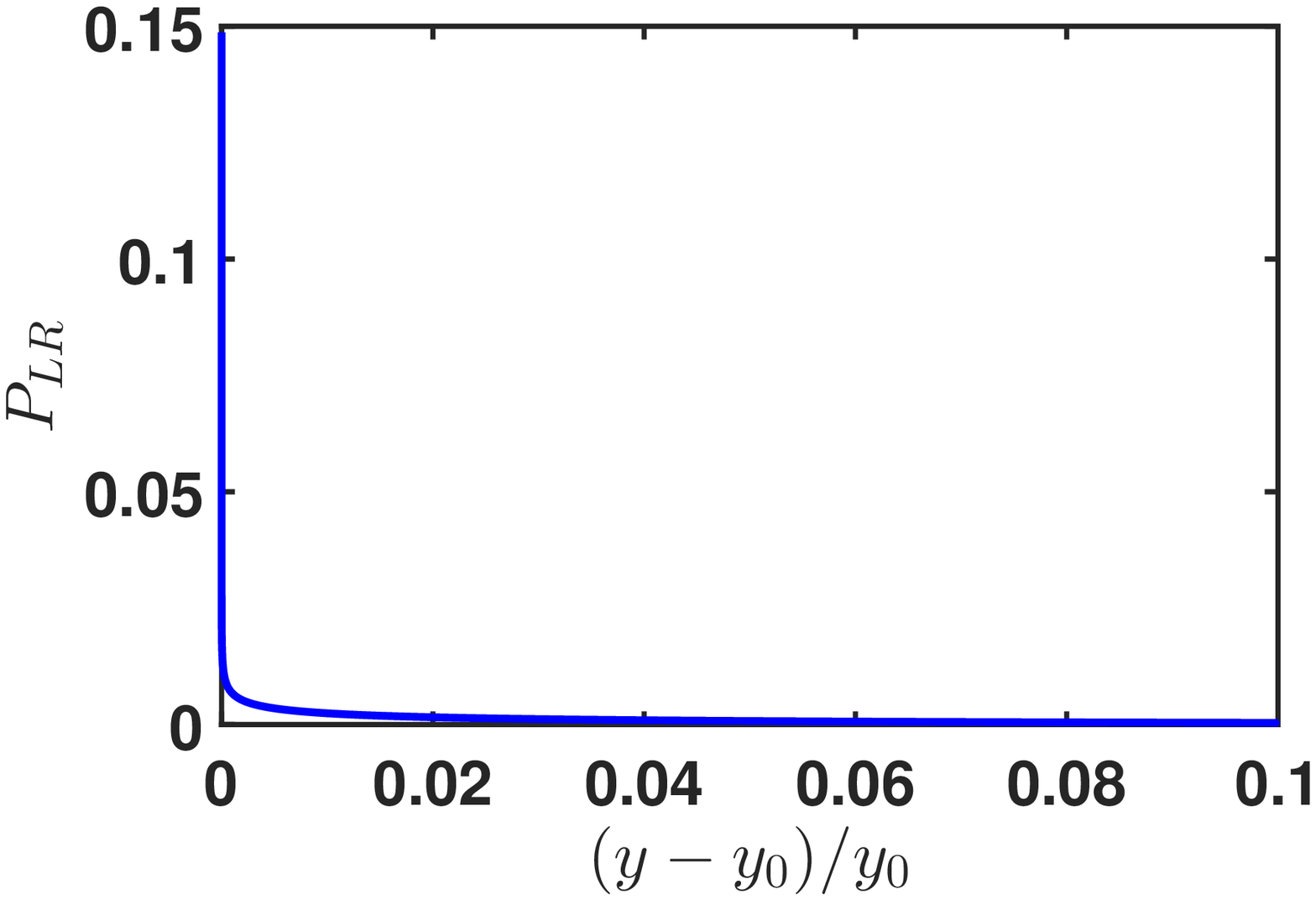}}
  \hskip-.6cm
  \subfigure[]
  {\label{2b}
  \includegraphics[scale=.4]{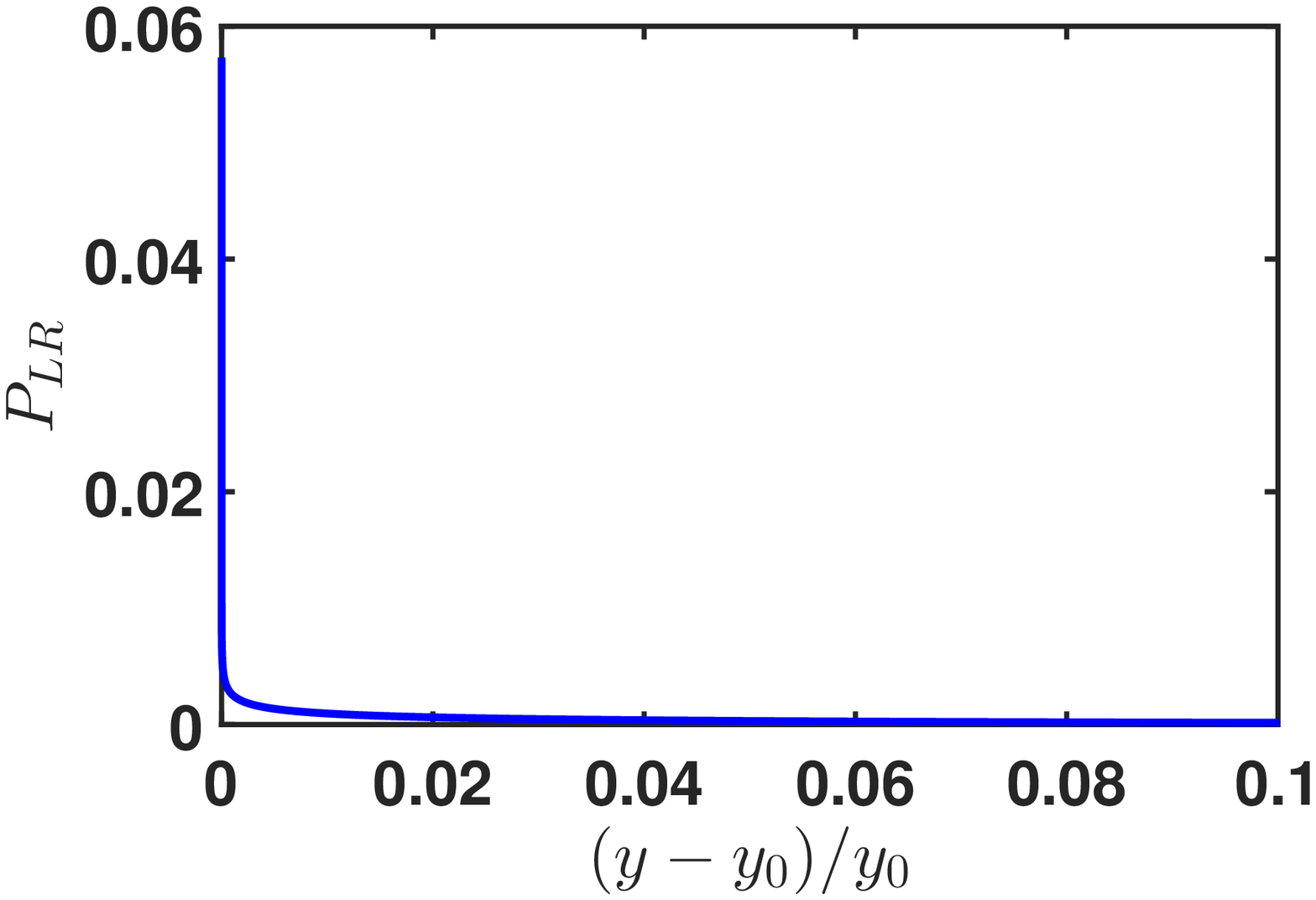}}
  \\
  \subfigure[]
  {\label{2c}
  \includegraphics[scale=.4]{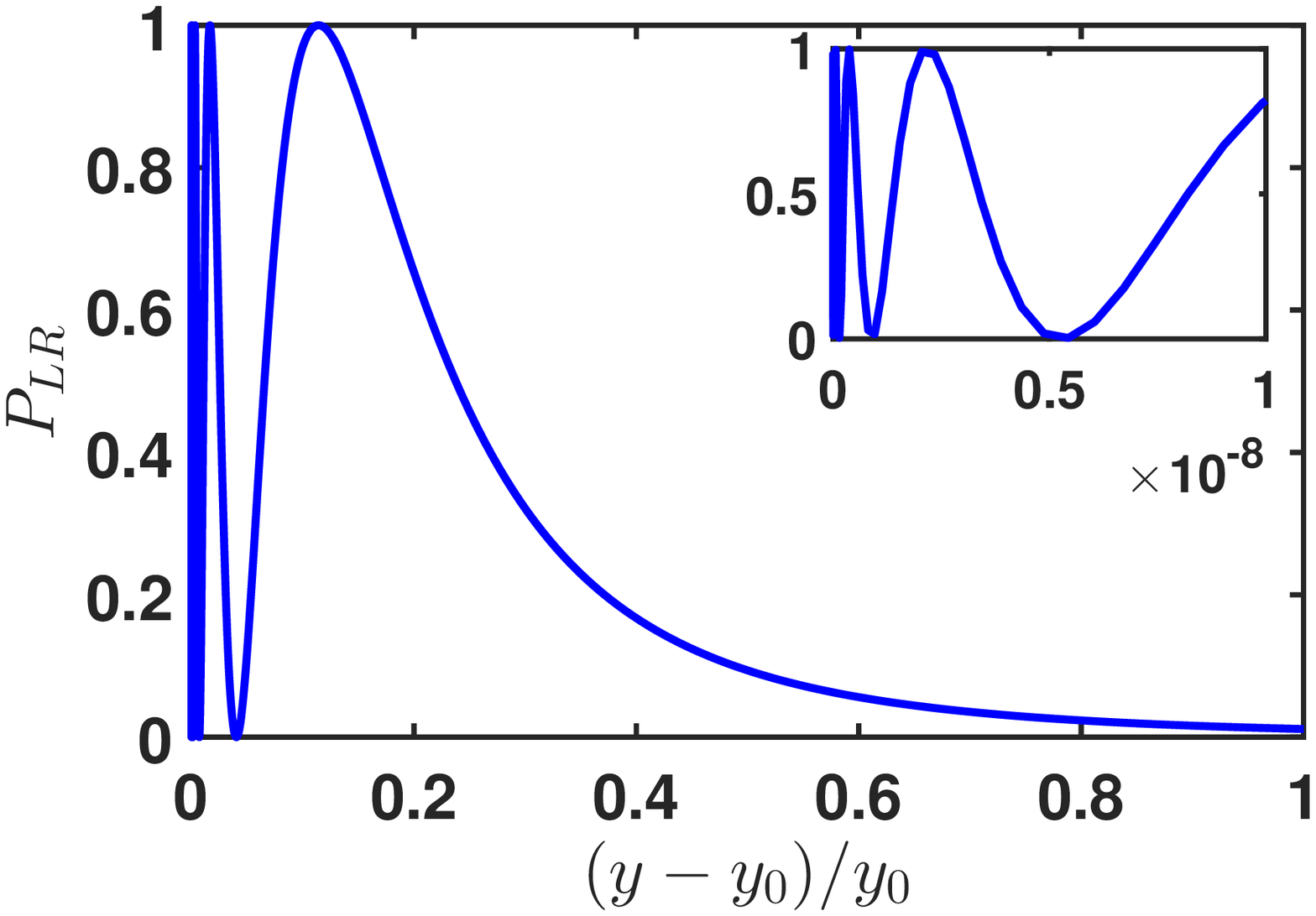}}
  \hskip-.6cm
  \subfigure[]
  {\label{2d}
  \includegraphics[scale=.4]{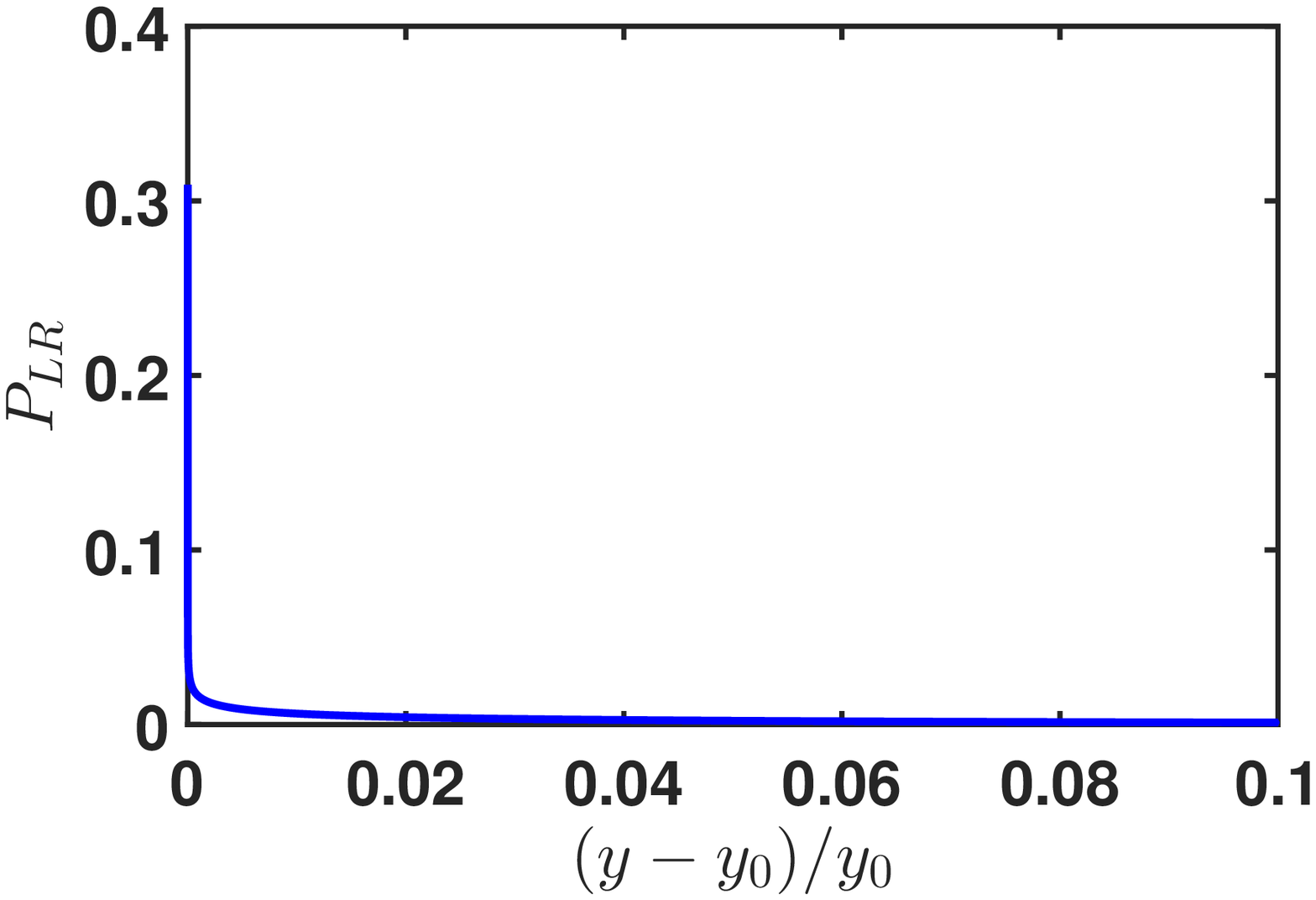}}
  \protect
  \caption{The transition probability $P_\mathrm{LR}$
  of spin oscillations versus the dimensioless impact parameter $y$
  for ultrarelativistic neutrinos scattering off a Kerr BH for different angular momenta $J$ of BH.
  (a)~and (c)~correspond to the direct scattering; (b)~and (d)~to the retrograde scattering.
  (a)~and (b)~$J = 0.1 M r_g$; (c)~and (d)~$J = 0.5 M r_g$.  
  \label{fig:PLRKerr}}
\end{figure}

The transition  probability of spin oscillations $P_\mathrm{LR}=(1 + \cos\alpha_{+\infty})/2$, computed on the basis of Eqs.~\eqref{eq:alphaintKerr}, \eqref{eq:FKultrarel}, and~\eqref{eq:RKultrarel} turns out to be nonzero. It means that there is a possibility of transitions between left and right polarized ultrarelarivistic neutrinos interacting with a rotating BH. The corresponding transition probabilities are shown in Fig.~\ref{fig:PLRKerr}.

Putting $z = 0$ in Eqs.~\eqref{eq:alphaintKerr}, \eqref{eq:FKultrarel}, and~\eqref{eq:RKultrarel}, one reproduces Eq.~\eqref{eq:abiggamma}. Thus, using the results of Sec.~\ref{sec:SCHWARZ}, one gets that the transition probability of spin oscillations is vanishing at a small angular momentum of BH. The same feature results from the comparison of Figs.~\ref{2a} and~\ref{2b} with Figs.~\ref{2c} and~\ref{2d}.

\section{Neutrino gravitational scattering accounting for the matter interaction\label{sec:MATT}}

In this section, we formulate the neutrino spin evolution equation
in background matter under the influence of a gravitational field
when a neutrino scatters off BH. Then, we derive the effective Schr\"{o}dinger
equation for spin oscillations of scattered neutrinos.

Using the forward scattering approximation, one gets that the neutrino
interaction with background matter is described by the following effective
Lagrangian in the Minkowski spacetime~\cite{MohPal04}:
\begin{equation}\label{eq:Largmat}
  \mathcal{L}_\mathrm{matt}=
  -\frac{G_{\mathrm{F}}}{\sqrt{2}}\bar{\nu}\gamma^{\mu}(1-\gamma^{5})\nu G_{\mu},
\end{equation}
where $\nu$ is the neutrino bispinor, $\gamma^{\mu}$ and $\gamma^{5}$
are the Dirac matrices, and $G_{\mathrm{F}}=1.17\times10^{-5}\,\text{GeV}^{-2}$
is the Fermi constant. The four vector $G^{\mu}$ is the linear combination
of the hydrodynamic currents and polarizations of background fermions.
It depends on the chemical composition of matter and the type of a
neutrino. The explicit form of $G^{\mu}$ can be found in Ref.~\cite{DvoStu02}.

Basing on Eq.~\eqref{eq:Largmat}, the influence of the neutrino interaction
with background matter on its spin evolution in curved spacetime was
studied in Refs.~\cite{Dvo13,Dvo19}. It results in the appearance
of the additional components of the vector $\bm{\Omega}_{g}$ in Eq.~(\ref{eq:Omega2}):
$\bm{\Omega}_{g}\to\bm{\Omega}=\bm{\Omega}_{g}+\bm{\Omega}_\mathrm{matt}$, where the vector $\bm{\Omega}_\mathrm{matt}$ has the form,
\begin{equation}\label{eq:Omegam}
  \bm{\Omega}_\mathrm{matt}=
  \frac{G_{\mathrm{F}}}{\sqrt{2}U^{t}}
  \left[
    \mathbf{u}
    \left(
      g^{0} - \frac{(\mathbf{gu})}{1+u^0}
    \right)
    - \mathbf{g}
  \right],
\end{equation}
where $g^a = (g^0,\mathbf{g}) = e^a_{\,\mu} G^\mu$ is the four vector of the effective neutrino interaction with background matter in the locally Minkowskian frame with coordinates $x^a$, $e^a_{\,\mu} = \partial x^a/\partial x^\mu$ are the vierbein vectors, and  $G^\mu$ is the analogue of the four vector in Eq.~\eqref{eq:Largmat} given in curved spacetime.

If we study the neutrino interaction with nonrelativistic unpolarized matter, only $G^0 = n_\mathrm{eff} U^{t}_f \neq 0$, where $U^{t}_f$ is the time component of the four velocity of plasma. For spin oscillations of electron neutrinos in the electrically
neutral hydrogen plasma one has $n_{\mathrm{eff}}=n_{e}$, where $n_{e}$
is the electron number density. The expressions for $n_{\mathrm{eff}}$
for other neutrino oscillations channels and various types of background
fermions can be found in Ref.~\cite{DvoStu02}.

Instead of dealing with Eq.~(\ref{eq:spinevgen}) for the spin precession,
it is convenient to study the neutrino polarization density matrix,
$\rho=\tfrac{1}{2}[1+(\bm{\sigma\zeta})]$, which obeys the equation,
$\mathrm{i}\dot{\rho}=[H,\rho]$, where $H=-(\bm{\sigma\Omega})$
and $\bm{\Omega}$ includes both the gravity and matter contributions
in Eqs.~(\ref{eq:Omega2}) or~\eqref{eq:Omega2Kerr}, and~(\ref{eq:Omegam}). Here $\bm{\sigma}$ are the Pauli matrices.

Since the Liouville\textendash von Neumann equation for the density
matrix is rather complicated for the analysis, we can use the Schr\"{o}dinger
equation, $\mathrm{i}\dot{\psi}=H\psi$. As we mentioned in Sec.~\ref{sec:GRAV},
neutrinos move along the first axis in the locally Minkowskian frame
at $r\to\infty$. Hence, it is convenient to use this axis for the
spin quantization. It mean that we should replace the Hamiltonian
$H\to \mathcal{U}_{2}H\mathcal{U}_{2}^{\dagger}$, where $\mathcal{U}_{2}=\exp(\mathrm{i}\pi\sigma_{2}/4)$.
This procedure brings the meaning to the effective wave function $\psi$. As in Secs.~\ref{sec:SCHWARZ} and~\ref{sec:KERR}, it is convenient to rewrite the Schr\"{o}dinger equation using the normalized radial coordinate $x=r/r_g$,
\begin{equation}\label{eq:Schr}
  \mathrm{i}\frac{\mathrm{d}\psi}{\mathrm{d}x}= H_{x}\psi,
  \quad
  H_{x}= - \mathcal{U}_{2}(\bm{\sigma\Omega}_{x})\mathcal{U}_{2}^{\dagger},
\end{equation}
where $\bm{\Omega}_{x} = r_g \bm{\Omega} \mathrm{d}t/\mathrm{d}r$. Now we are ready to  write down the effective Hamiltonian $H_{x}$ for both the Schwarzschild and Kerr metrics.

One has that $U^{t}_f=(1-r_g/r)^{-1}$ and $\mathbf{U}_f=0$ for nonrelativistic plasma near a nonrotating BH. We found in Ref.~\cite{Dvo06} that $e_{\,\mu}^{0}=(\sqrt{1-r_g/r},0,0,0)$
is the vierbein vector in the Schwarzschild metric. Then, in Eq.~\eqref{eq:Omegam}, $g^0 = n_\mathrm{eff} (1-r_g/r)^{-1/2}$ and $\mathbf{g}=0$.

Using Eqs.~\eqref{eq:uexpl} and~\eqref{eq:eqmtr}, we obtain all the components of the vector $\bm{\Omega}_x$ in the form,
\begin{equation}\label{eq:OmegaxSchw}
  \Omega_{x1} = \frac{V x}{x-1},
  \quad
  \Omega_{x2} = \pm\frac{1}{4}F_\mathrm{S}(x),
  \quad
  \Omega_{x3} = \pm\frac{Vxy}{\sqrt{R_\mathrm{S}(x)(x-1)}},
\end{equation}
where $V = G_{\mathrm{F}}n_{\mathrm{eff}} r_g / \sqrt{2}$ is the dimensionless effective potential, the functions $F_\mathrm{S}(x)$ and $R_\mathrm{S}(x)$ are given in Eqs.~\eqref{eq:aobslim} and~\eqref{eq:eqtosolve}, the upper and the lower signs stay for outgoing and incoming neutrinos.

In the case of a rotating BH, the four velocity of nonrelativistic matter has the components,
\begin{equation}
  U^{t}_f=\frac{r^{3}+a^{2}(r+r_{g})}{r[r(r-r_{g})+a^{2}]},
\end{equation}
and $\mathbf{U}_f=0$. The four vector $g^a$ was obtained in Ref.~\cite{Dvo13} in the form,
\begin{equation}\label{eq:ga}
  g^a = \frac{n_\mathrm{eff} U_f^t\sqrt{r}}{\sqrt{r^3+a^2(r+r_g)}}
  \left(
    \sqrt{r(r-r_g)+a^2},0,0,-\frac{ar_g}{r}
  \right),
\end{equation}
where we take again that matter is nonrelativistic.

Basing on Eqs.~\eqref{eq:u013}, \eqref{eq:ga}, and~\eqref{eq:eqmtrKerr}, we obtain the components of $\bm{\Omega}_x$ in the Kerr metric as
\begin{align}\label{eq:OmegaxKerr}
  \Omega_{x1} = & C V \frac{\sqrt{x[x^3 + z^2(x+1)]}}{x(x-1)+z^2},
  \quad
  \Omega_{x2} = \pm\frac{1}{4}F_\mathrm{K}(x),
  \notag
  \\
  \Omega_{x3} = & \pm\frac{V}{\sqrt{(1-\gamma^{-2})R_\mathrm{K}(x)[x(x-1)+z^2]}},
  \notag
  \\
  & \times
  \left[
    C x^{3/2}y\sqrt{1-\gamma^{-2}}+
    z\gamma^{-1}\frac{\sqrt{x^3+z^2(x+1)}}{\sqrt{x(x-1)+z^2}}
  \right],
  \notag
  \\
  C = & \frac{x^3 + z^2(x+1)+\gamma^{-1}\sqrt{x[x(x-1)+z^2][x^3+z^2(x+1)]}}
  {x^3 + z^2(x+1)-zy\sqrt{1-\gamma^{-2}}+\gamma^{-1}\sqrt{x[x(x-1)+z^2][x^3+z^2(x+1)]}},
\end{align}
where the functions $F_\mathrm{K}(x)$ and $R_\mathrm{K}(x)$ are given in Eqs.~\eqref{eq:alphaKerr} and~\eqref{eq:RKerr}.

We revealed in Sec.~\ref{sec:KERR} that there is a spin conversion of ultrarelativistic neutrinos scattered off a Kerr BH. Thus, it is reasonable to rewrite Eq.~\eqref{eq:OmegaxKerr} for such particles. We get that
\begin{align}\label{eq:OmegaxKerrultrarel}
  \Omega_{x1} = &  \frac{V\sqrt{x}[x^3 + z^2(x+1)]^{3/2}}{[x^3 + z^2(x+1)-zy][x(x-1)+z^2]},
  \notag
  \\
  \Omega_{x3} = & \pm \frac{V y x^{3/2}[x^3 + z^2(x+1)]}
  {[x^3 + z^2(x+1)-zy]\sqrt{R_\mathrm{K}(x)[x(x-1)+z^2]}},
\end{align}
when $\gamma\gg 1$. The expression for $\Omega_{x2}$ straightforwardly results from Eq.~\eqref{eq:FKultrarel}. In Eq.~\eqref{eq:OmegaxKerrultrarel}, the function $R_\mathrm{K}(x)$ is given by Eq.~\eqref{eq:RKultrarel}.

Equation~(\ref{eq:Schr}) should be supplied with the initial condition
$\psi_{-\infty}^{\mathrm{T}}=(1,0)$, which means that all incoming
neutrinos are left polarized. Since the neutrino velocity $\mathbf{u}$ changes
the direction at $t\to+\infty$, the transition probability reads
$P_{\mathrm{LR}}=|\psi_{+\infty}^{(1)}|^{2}$, and, correspondingly,
the survival probability is $P_{\mathrm{LL}}=|\psi_{+\infty}^{(2)}|^{2}$,
where $\psi_{+\infty}^{\mathrm{T}}=(\psi_{+\infty}^{(1)},\psi_{+\infty}^{(2)})$
is the asymptotic solution of Eq.~(\ref{eq:Schr}).

The solution of Eq.~(\ref{eq:Schr}), with $\bm{\Omega}_x$ in Eqs.~(\ref{eq:OmegaxSchw}) or~(\ref{eq:OmegaxKerr}), can
be found only numerically because of the nontrivial dependence of
$\bm{\Omega}_{x}$ on $x$. Moreover, in Sec.~\ref{sec:APPL}, we
discuss the situation when $n_{\mathrm{eff}}=n_{\mathrm{eff}}(r)$, or $V=V(x)$,
which makes the analysis more complicated.

We also mention, that we cannot integrate Eqs.~(\ref{eq:Schr})
to the turn point $x_{m}$ and then automatically reconstruct $\psi_{+\infty}$,
as we made in Secs.~\ref{sec:SCHWARZ} and~\ref{sec:KERR} to find $\alpha_{+\infty}$.
In the presence of the background matter, the neutrino spin precesses around the axis with nonconstant direction. Moreover, the components $\Omega_{x2,3}$ in Eqs.~\eqref{eq:OmegaxSchw} and~\eqref{eq:OmegaxKerr} change the sign at $x=x_{m}$.
Thus, to obtain $\psi_{+\infty}$, one should integrate Eq.~(\ref{eq:Schr}), first, in the interval $+\infty>x>x_{m}$
and, then, for $x_{m}<x<+\infty$, with the solutions being stitched
at $x_{m}$. While integrating Eq.~(\ref{eq:Schr}) in these two intervals, one should account for the signs of $\Omega_{x2,3}$. This fact significantly reduces the accuracy of the numerical
simulation compared to Secs.~\ref{sec:SCHWARZ} and~\ref{sec:KERR}.

\section{Astrophysical applications\label{sec:APPL}}

In this section, we present the numerical solutions of Eq.~(\ref{eq:Schr}) for the neutrino scattering off SMBH surrounded by an accretion disk. We discuss both rotating and nonrotating SMBH, as well as the cases of ultrarelativistic neutrinos and neutrinos having a finite energy. The measurable neutrino fluxes are obtained.

First we notice, that standard model neutrinos are produced as left
polarized particles. If they gravitationally interact with BH, some
incoming left neutrinos become right polarized after scattering. A
neutrino detector can observe only left neutrinos. Hence, the observed
flux of neutrinos is $F_{\nu}=P_{\mathrm{LL}}F_{0}$, where $F_{0}$
is the flux of scalar particles. The value of $F_{0}$ is proportional
to the differential cross section, $F_{0}\sim\mathrm{d}\sigma/\mathrm{d}\varOmega$,
which is studied in Appendices~\ref{sec:PARTM} and~\ref{sec:PARTMKERR}.

We assume that the neutrino beam scatters off a SMBH surrounded by
an accretion disk. For example, we can suppose that such a SMBH is
in the center of a Seyfert galaxy. We take that the plasma density
in the disk scales as $n_{e}\propto r^{-\beta}$. The value of $\beta$
is very model dependent. For example, $\beta\approx0.5$ in an advection
dominated accretion disk studied in Ref.~\cite{Igu00}. If we take
that the mass of SMBH in question is $M\sim10^{8}M_{\odot}$, the
plasma density in the vicinity of SMBH can be up to $n_{e}\sim10^{18}\,\text{cm}^{-3}$~\cite{Jia19}. Thus, the dimensionless effective potential $V(r)=G_\mathrm{F}n_{e}(r)r_{g}/\sqrt{2}$,
reads $V(x)=V_{\mathrm{max}}x^{-\beta}$, where $x=r/r_{g}$.

The plasma motion in an accretion disk is driven not only by the gravitational interaction with a central BH. It also depends on the interaction between charged particles. Thus, the plasma angular velocity can be a complicated function of $r$, which is model dependent. We omit this additional factor in the description of the neutrino spin evolution and assume that the plasma motion in an accretion disk is nonrelativistic.

\begin{figure}
  \centering
  \includegraphics[scale=.4]{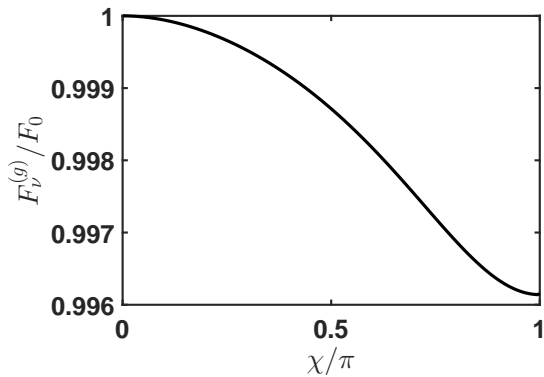}
  \protect
  \caption{The ratio of fluxes of massive neutrinos, obtained using Eq.~\eqref{eq:aobslim},
  and scalar particles in their  
  scattering off a Schwarzschild BH versus the scattering angle $\chi$ normalized
  by $\pi$.\label{fig:FgF0Schwarz}}
\end{figure}

First, in Fig.~\ref{fig:FgF0Schwarz}, we show the ratio of the fluxes of massive neutrinos, $F_\nu^{(g)}$, scattered off a nonrotating BH and scalar particles $F_0$ of the same mass, with spin oscillations of neutrinos in the gravitational field being accounted for. These fluxes are proportional to the differential cross
section, $F\sim\mathrm{d}\sigma/\mathrm{d}\varOmega$, where $\mathrm{d}\varOmega=2\pi\sin\chi\mathrm{d}\chi$. The calculation of the cross section for scalar particles is presented in Appendix~\ref{sec:PARTM} and the result is shown in Fig.~\ref{fig:diffsc0}. We do not take into account the neutrino interaction with matter in Fig.~\ref{fig:FgF0Schwarz}. 

The neutrino energy is $E=10m$ in Fig.~\ref{fig:FgF0Schwarz}. We have seen in Sec.~\ref{sec:SCHWARZ} that neutrino spin oscillations in the scattering in the Schwarzschild metric are vanishing if $E\gg m$. Thus, the further enhancement of the neutrino energy will result in $F_\nu^{(g)}$ practically coinciding with $F_0$. Consideration of neutrino energies smaller than in  Fig.~\ref{fig:FgF0Schwarz} is inexpedient from the point of view of possible astrophysical applications accounting for the current upper bound on neutrino masses in Ref.~\cite{Ake19}. 

One can see in Fig.~\ref{fig:FgF0Schwarz} that $F_\nu^{(g)}<F_0$. This fact is owing to the survival probability $P_\mathrm{LL}<1$, which $F_\nu^{(g)}$ is proportional to, for the gravitational scattering of massive neutrinos. The maximal difference between the fluxes is for the backward neutrino scattering at $\chi=\pi$. However, the maximal deviation of $F_\nu^{(g)}$ from $F_0$ is less than 1\%. Accounting for the rather small neutrino energy, it makes difficult to observe the effect of spin oscillations in the neutrino scattering off a nonrotating BH. 

\begin{figure}
  \centering
  \subfigure[]
  {\label{4a}
  \includegraphics[scale=.4]{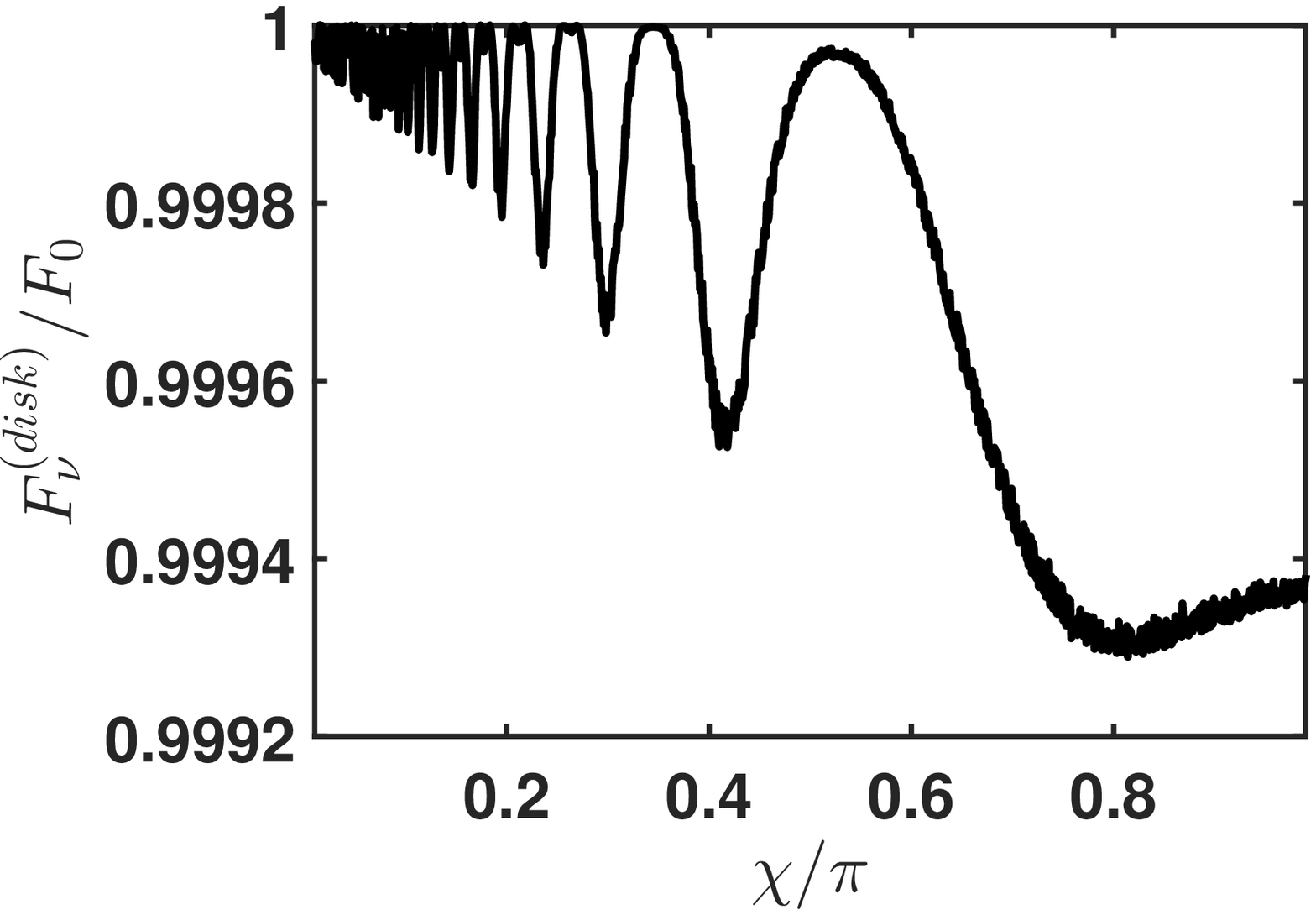}}
  \hskip-.6cm
  \subfigure[]
  {\label{4b}
  \includegraphics[scale=.4]{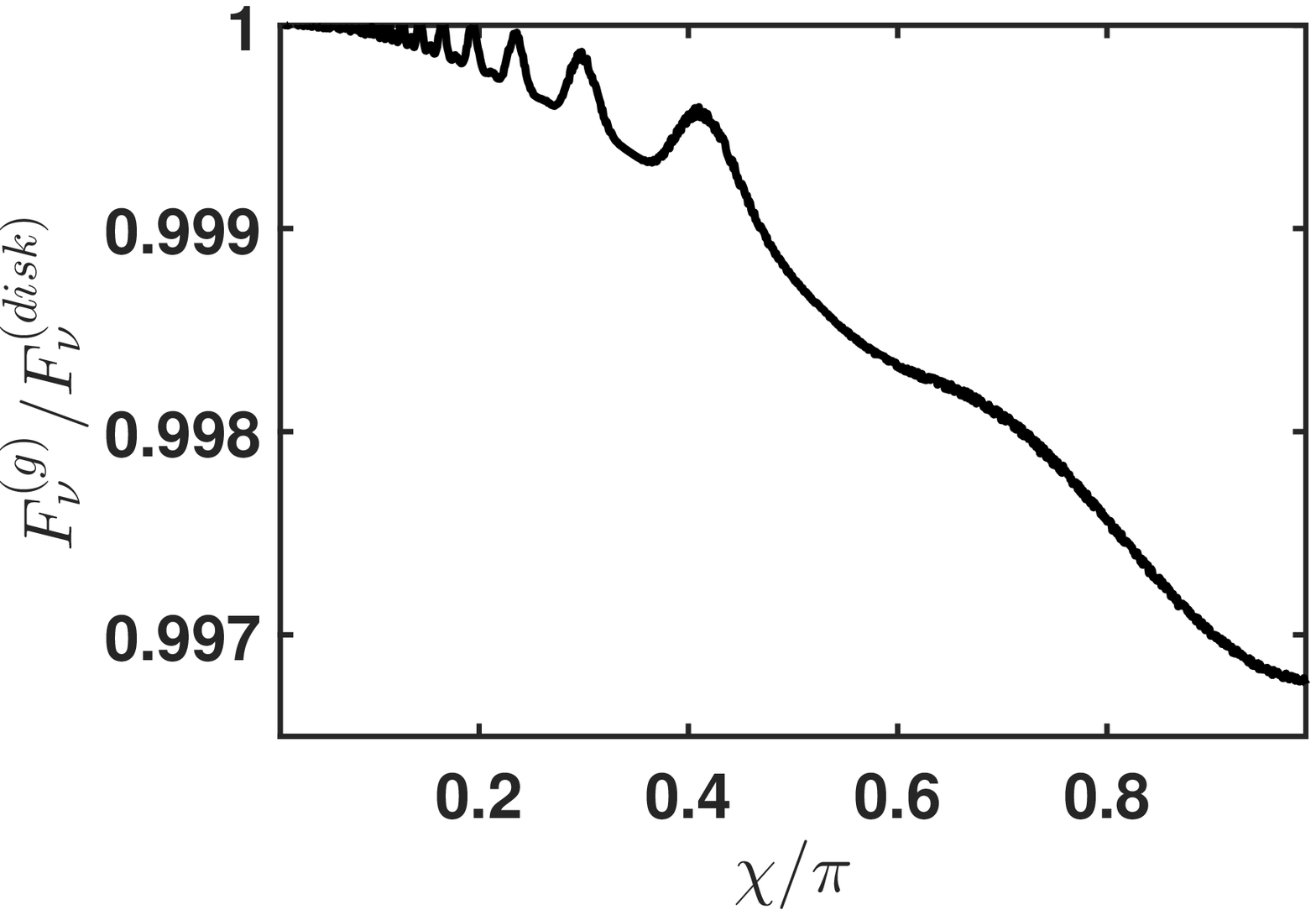}}
  \protect
  \caption{The ratios of fluxes obtained by the numerical solution of 
  Eqs.~\eqref{eq:Schr} and~\eqref{eq:OmegaxSchw}
  versus $\chi/\pi$.
  (a)~The flux of scattered neutrinos, accounting for the matter interaction,
  $F_\nu^{(\mathrm{disk})}$
  normalized by $F_0$;
  (b)~the ratio of $F_\nu^{(g)}$ and $F_\nu^{(\mathrm{disk})}$.
  The energy of particles $E=10m$.\label{fig:Schwarzmatt}}
\end{figure}

Now we study the influence of the interaction with an accretion disk matter on the flux of neutrinos scattered off a nonrotating SMBH with $M= 10^8 M_\odot$. In Fig.~\ref{fig:Schwarzmatt}, we show the result of the numerical solution of Eqs.~\eqref{eq:Schr} and~\eqref{eq:OmegaxSchw}. We take that $n_e^{(\mathrm{max})} = 2\times10^{18}\,\text{cm}^{-3}$ at $r=r_g$ and $\beta = 0.2$. The flux of scattered neutrinos with $E=10m$ accounting for the matter interaction $F_\nu^{(\mathrm{disk})}$ , normalized by the flux of scalar particles, is shown in Fig.~\ref{4a}. The ratio of  $F_\nu^{(g)}$ and $F_\nu^{(\mathrm{disk})}$ is presented in Fig.~\ref{4b}.

One can see in Figs.~\ref{fig:FgF0Schwarz} and~\ref{fig:Schwarzmatt} that $F_\nu^{(g)}<F_\nu^{(\mathrm{disk})}<F_0$. The former inequality results from the fact that the matter interaction makes neutrino spin oscillations to be out of the resonance. Hence $P_\mathrm{LL}^{(g)}<P_\mathrm{LL}^{(\mathrm{disk})}$. The latter inequality is a consequence of $P_\mathrm{LL}^{(\mathrm{disk})}<1$. As in Fig.~\ref{fig:FgF0Schwarz}, the maximal deviations of the neutrino fluxes from $F_0$ in Fig.~\ref{fig:Schwarzmatt} are at $\chi=\pi$, Nevertheless, taking into account the magnitude of these deviations, one concludes that neutrino spin oscillations are negligible in the gravitational scattering off a nonrotating BH.

Now we turn to the consideration of spin oscillations of neutrinos scattered off a rotating BH. We have revealed in Sec.~\ref{sec:KERR} that ultrarelativistic neutrinos can change their polarization in this situation. Moreover, using Eqs.~\eqref{eq:OmegaxSchw} and~\eqref{eq:OmegaxKerr}, one gets that the effective Hamiltonian for spin oscillations in the Kerr metric coincides with that for the Schwarzschild metric at $z\to 0$. Therefore, to highlight the manifestation of spin effects in the gravitational scattering off a Kerr BH, we consider ultrarelativistic neutrinos and a maximally rotating BH with $z=1/2$ or $a=M$.

\begin{figure}
  \centering
  \subfigure[]
  {\label{5a}
  \includegraphics[scale=.4]{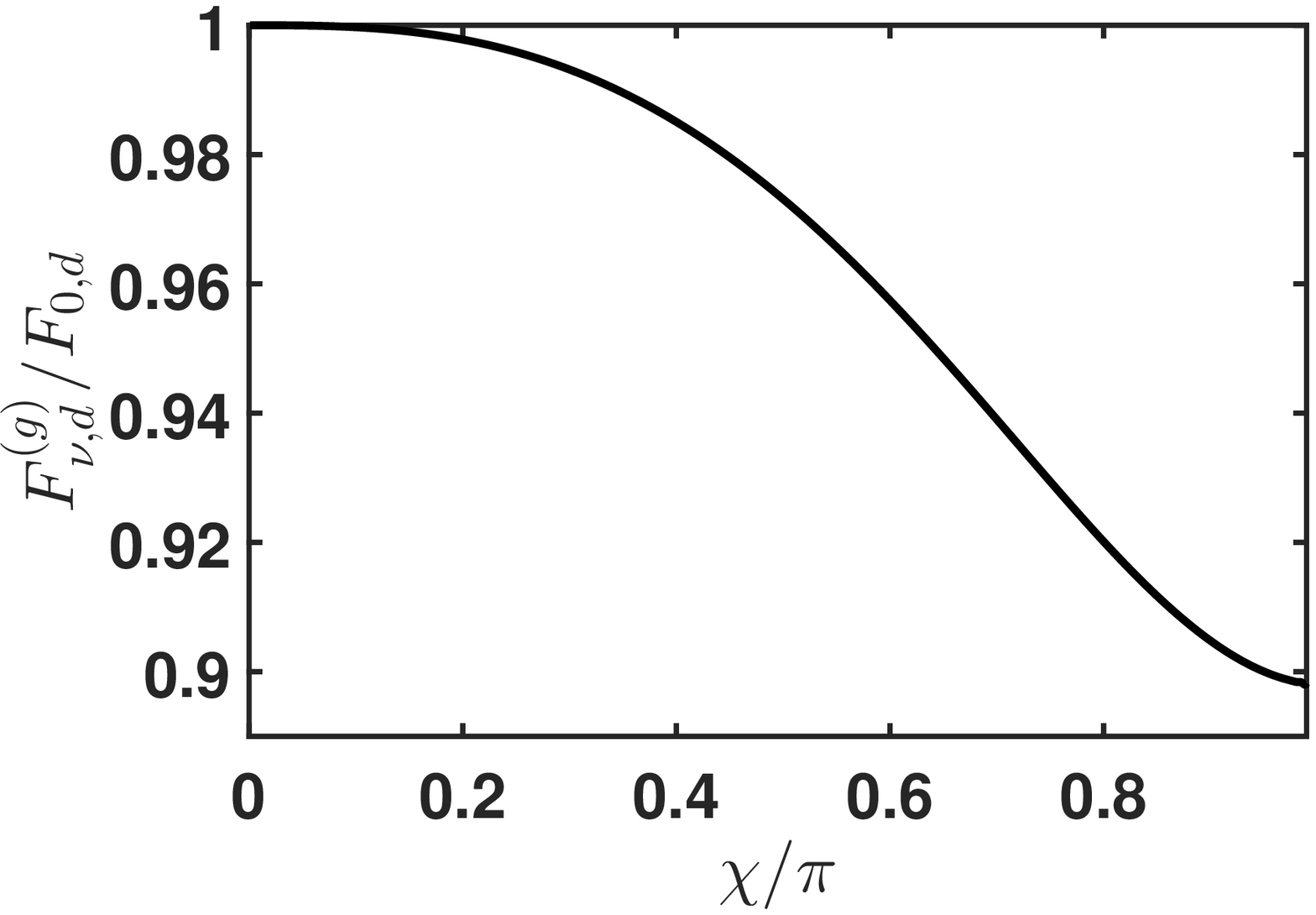}}
  \hskip-.6cm
  \subfigure[]
  {\label{5b}
  \includegraphics[scale=.4]{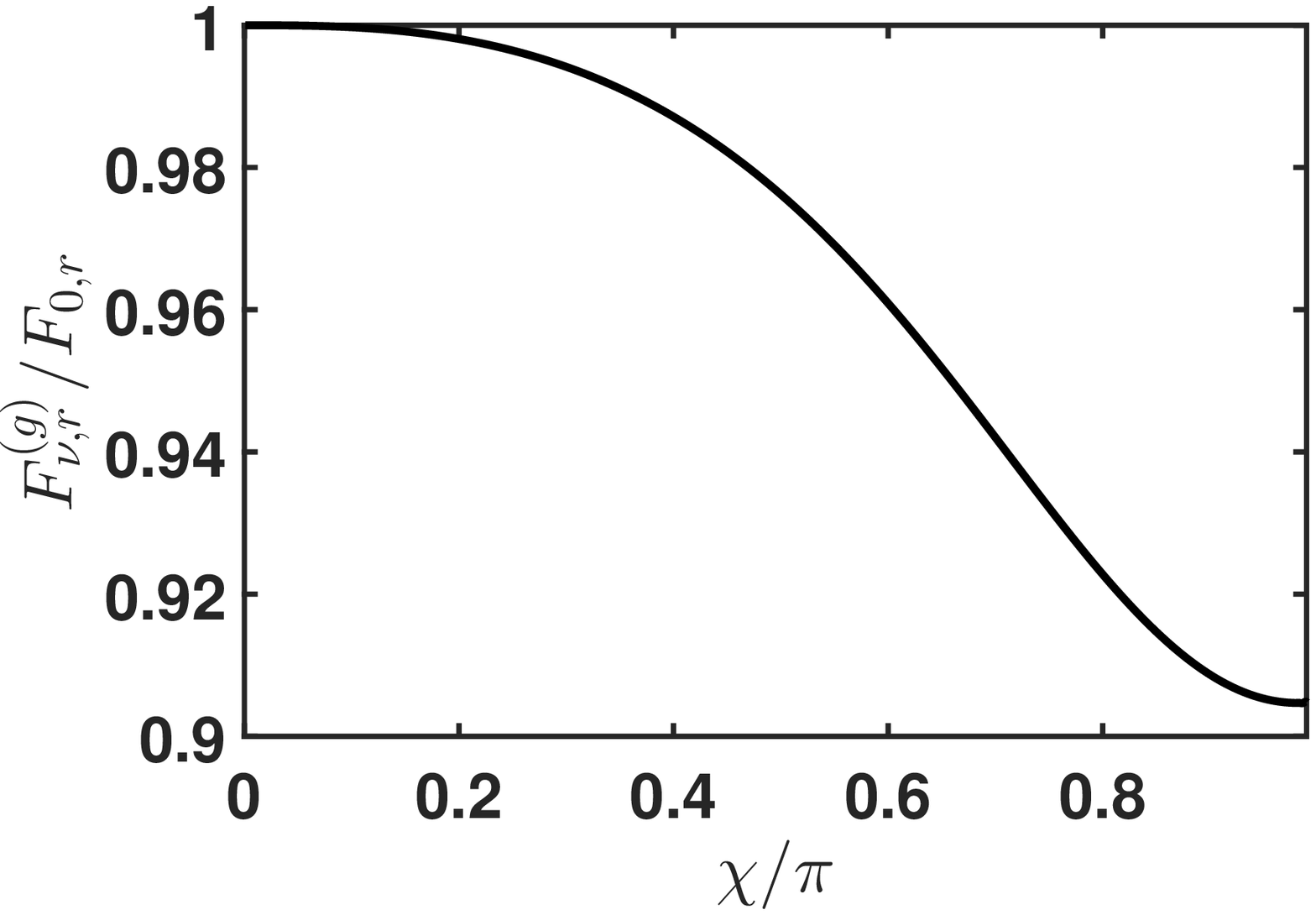}}
  \protect
  \caption{The ratios of fluxes of ultrarelativistic neutrinos, based on
  Eqs.~\eqref{eq:alphaintKerr} and~\eqref{eq:FKultrarel}, and scalar particles in their  
  scattering off a Kerr BH versus $\chi/\pi$.
  (a)~Direct scattering; 
  (b)~retrograde scattering.
  Particles are ultrarelativistic, having $E \gg m$. BH is maximally rotating with $a=M$.
  \label{fig:FgF0Kerr}}
\end{figure}

In Fig.~\ref{fig:FgF0Kerr}, we show the ratios of the fluxes of gravitationally scattered ultrarelativistic neutrinos $F_{\nu d,r}^{(g)}$ and scalar particles $F_{0 d,r}$ in the case of a rotating BH. We present the case of the direct scattering in Fig.~\ref{5a} and the retrograde one in Fig.~\ref{5b}. As for a Schwarzschild BH, $F_{\nu d,r}^{(g)} = F_{0 d,r}$ at $\chi=0$. However, $F_{\nu d,r}^{(g)}$ is about 10\% less than $F_{0 d,r}$ for backwardly scattered neutrinos with $\chi=\pi$.

\begin{figure}
  \centering
  \subfigure[]
  {\label{6a}
  \includegraphics[scale=.4]{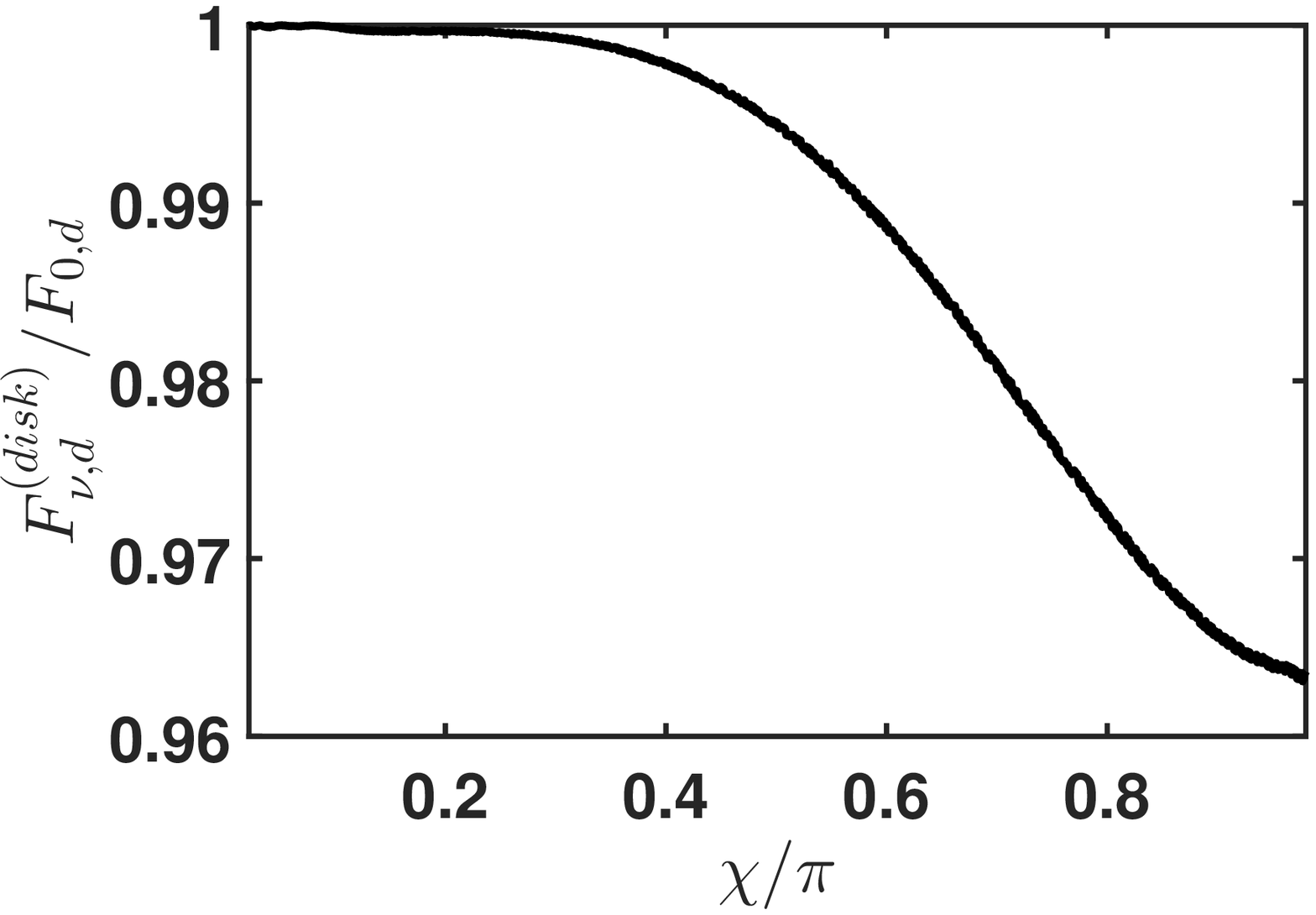}}
  \hskip-.6cm
  \subfigure[]
  {\label{6b}
  \includegraphics[scale=.4]{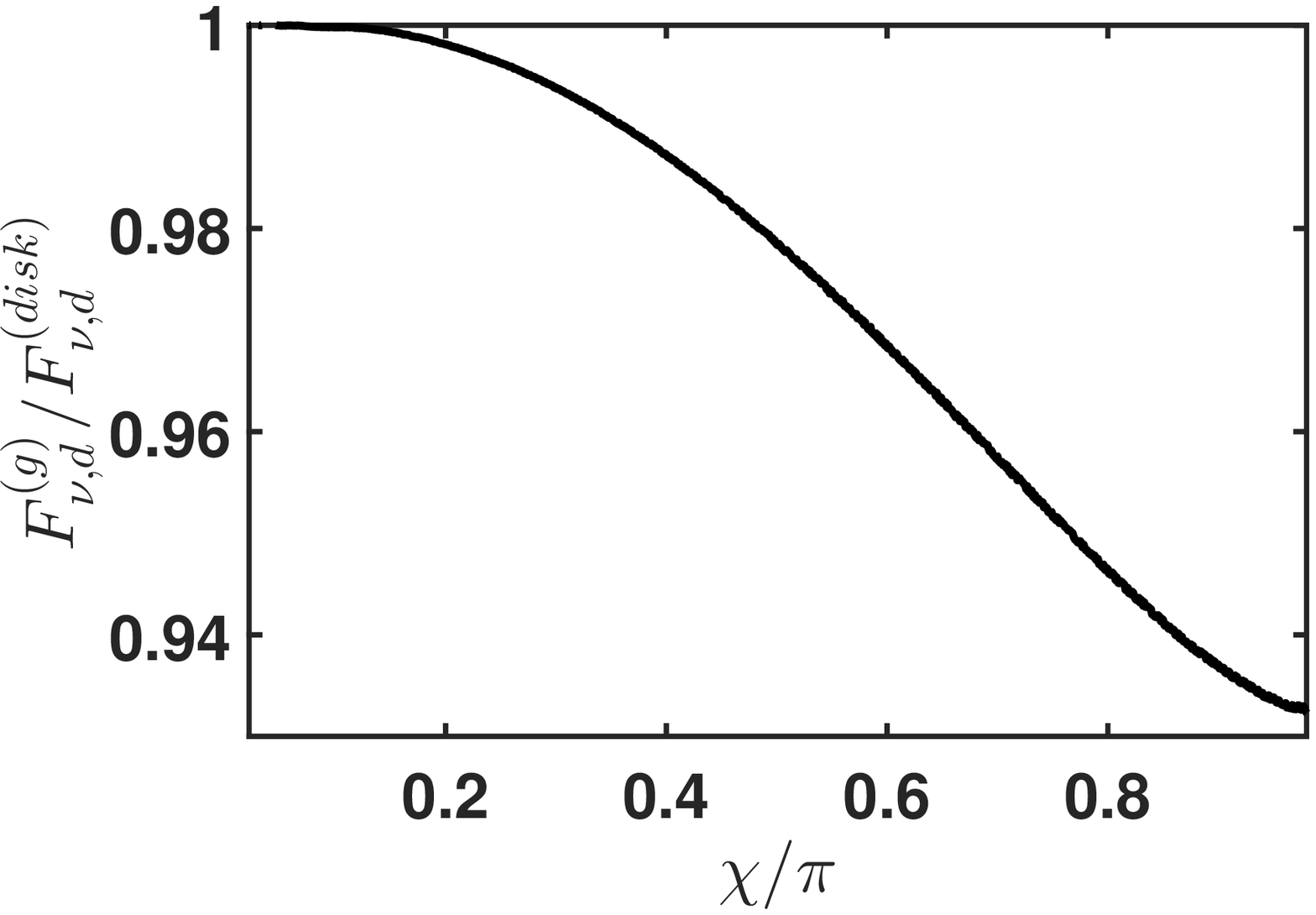}}
  \\
  \subfigure[]
  {\label{6c}
  \includegraphics[scale=.4]{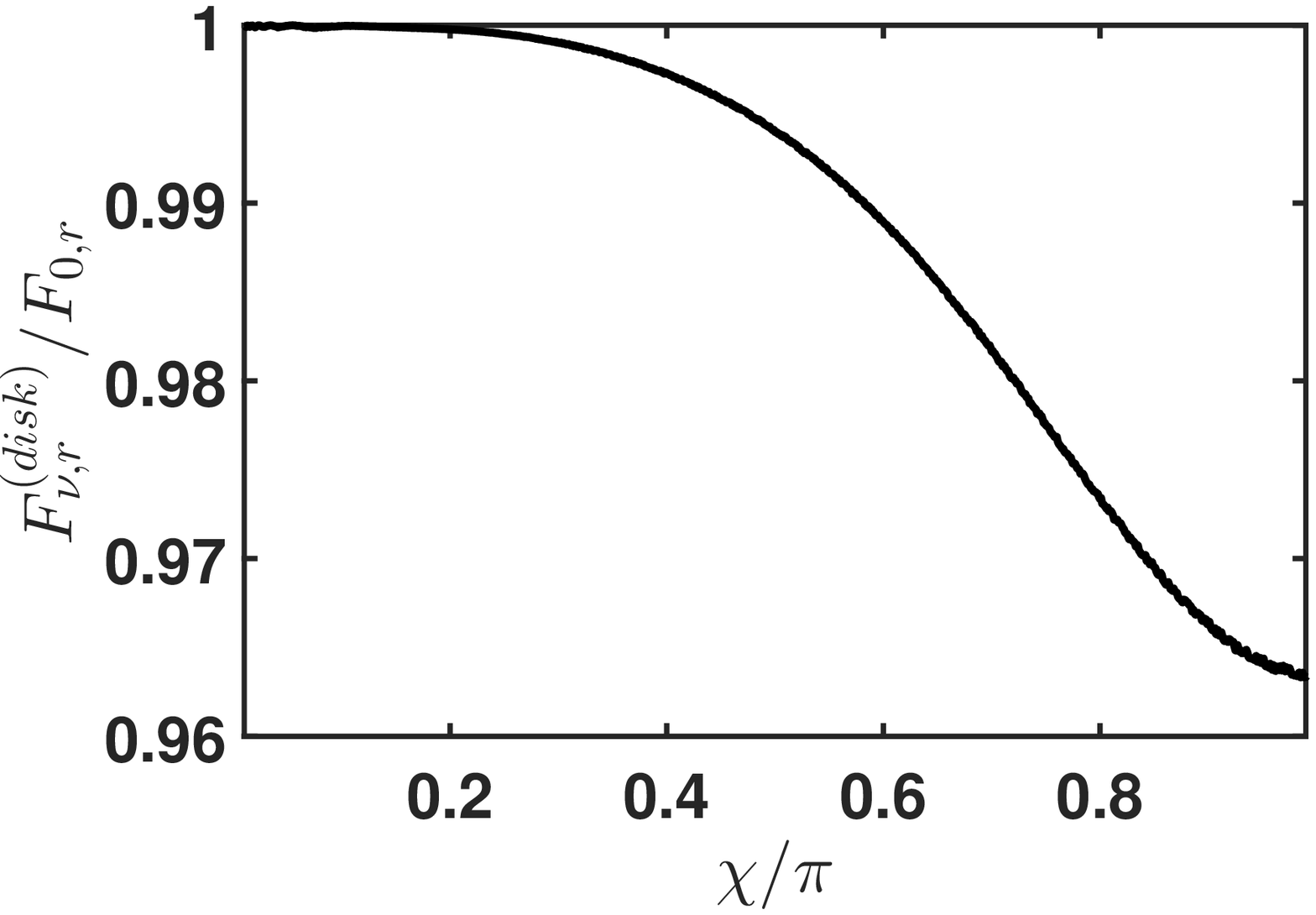}}
  \hskip-.6cm
  \subfigure[]
  {\label{6d}
  \includegraphics[scale=.4]{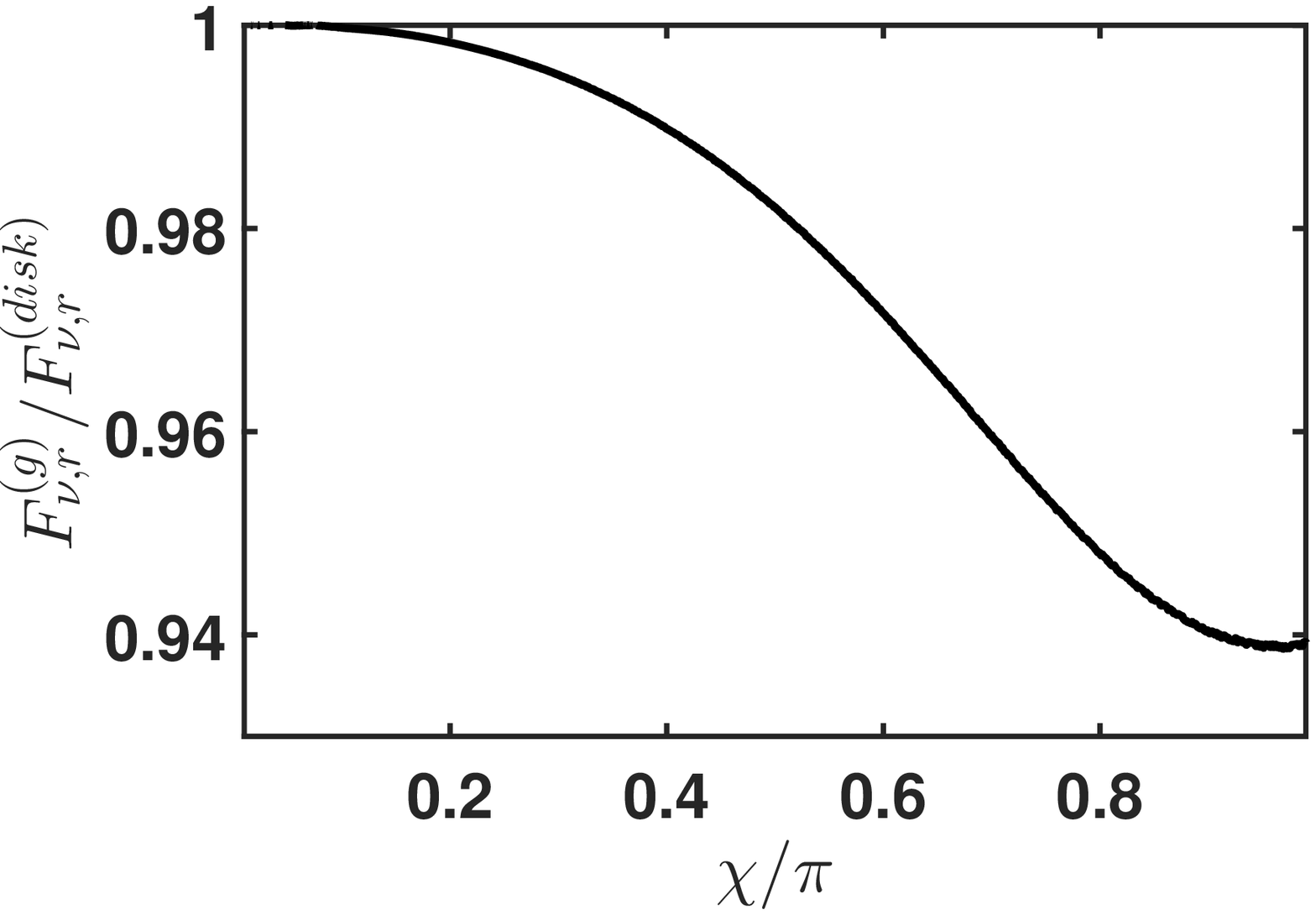}}
  \protect
  \caption{The ratios of fluxes obtained by the numerical solution of 
  Eqs.~\eqref{eq:Schr} and~\eqref{eq:OmegaxKerrultrarel}
  versus $\chi/\pi$.
  (a) and (b)~Direct scattering; 
  (c) and (d)~retrograde scattering.
  (a) and (c)~The fluxes of scattered neutrinos, accounting for the matter interaction,
  $F_{\nu d,r}^{(\mathrm{disk})}$
  normalized by $F_{0 d,r}$;
  (b) and (d)~the ratios of $F_{\nu d,r}^{(g)}$ and $F_{\nu d,r}^{(\mathrm{disk})}$.
  \label{fig:Kerrmatt}}
\end{figure}

Now we take into account the neutrino interaction with an accretion disk. We consider a situation of a maximally rotating SMBH with $M= 10^8 M_\odot$ surrounded by an accretion disk with $n_e^{(\mathrm{max})} = 10^{18}\,\text{cm}^{-3}$ at $r=r_g$ and $\beta = 0.5$. In Fig.~\ref{fig:Kerrmatt}, we depict the fluxes of ultrarelativistic neutrinos $F_{\nu d,r}^{(\mathrm{disk})}$, accounting for the interaction with background matter normalized by $F_{0 d,r}$, as well as the ratios of $F_{\nu d,r}^{(g)}$ and $F_{\nu d,r}^{(\mathrm{disk})}$. We present the cases of the direct scattering in Figs.~\ref{6a} and~\ref{6b}, and the retrograde one in Figs.~\ref{6c} and~\ref{6d}.

We can see in Fig.~\ref{fig:Kerrmatt} that the fluxes $F_{\nu d,r}^{(\mathrm{disk})}$ and $F_{0 d,r}$, as well as $F_{\nu d,r}^{(g)}$ and $F_{\nu d,r}^{(\mathrm{disk})}$, differ by about 5\%. The maximal difference is at $\chi=\pi$. The inequality $F_\nu^{(g)}<F_\nu^{(\mathrm{disk})}<F_0$, established for the Schwarzschild metric above, remains valid for a Kerr BH as well.

\begin{figure}
  \centering
  \subfigure[]
  {\label{7a}
  \includegraphics[scale=.4]{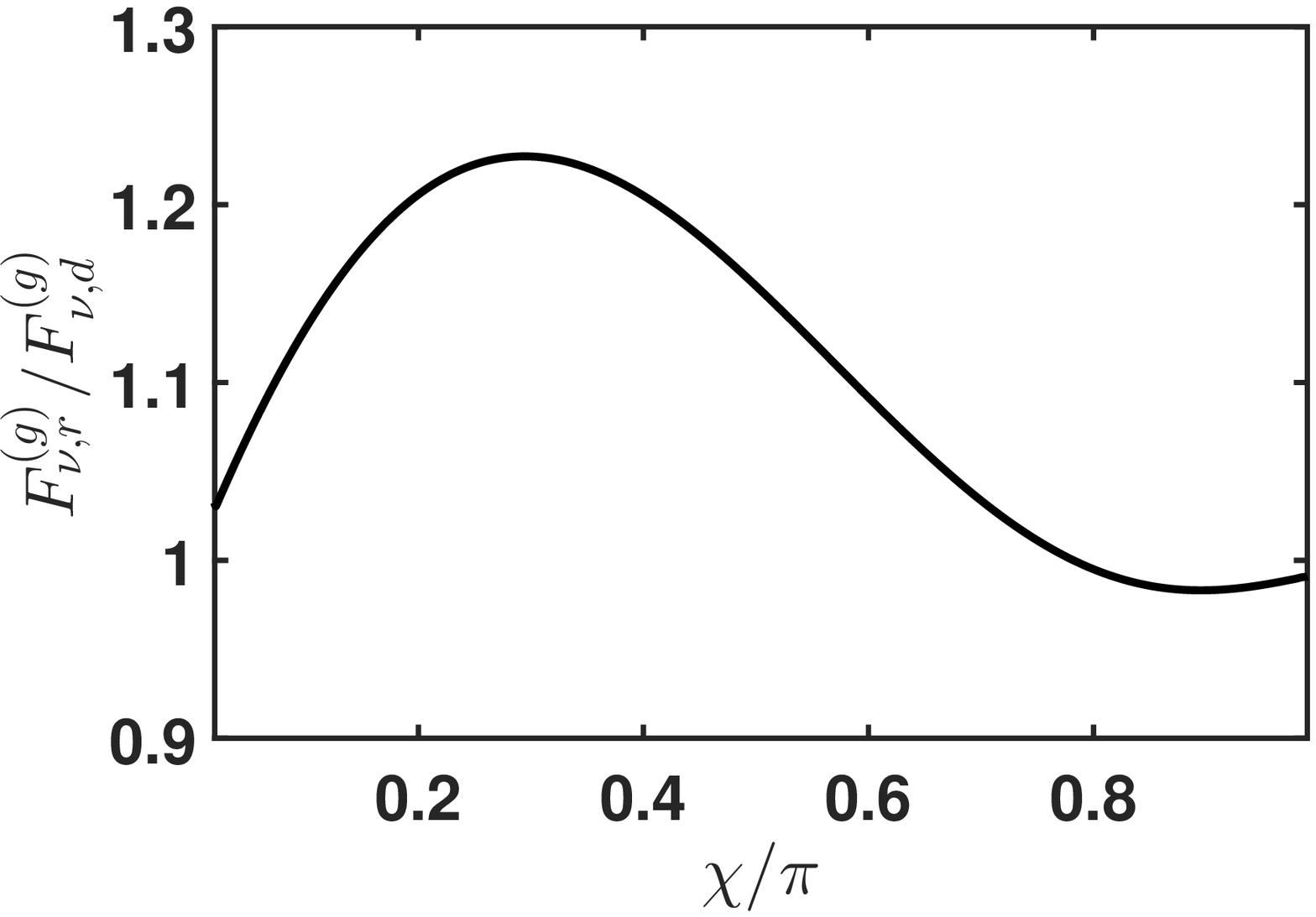}}
  \hskip-.6cm
  \subfigure[]
  {\label{7b}
  \includegraphics[scale=.4]{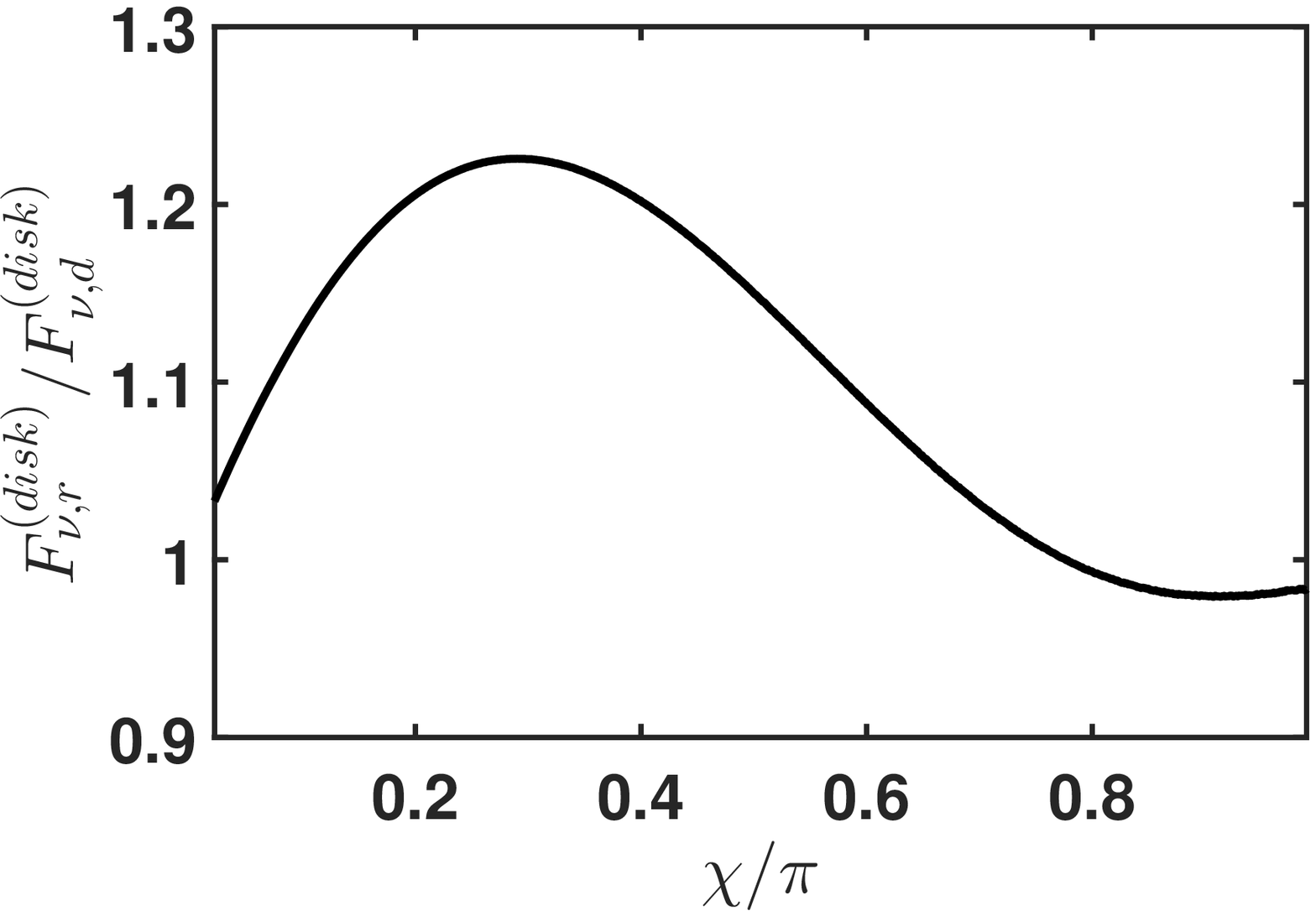}}
  \protect
  \caption{The ratios of fluxes for the direct and retrograde scatterings
  versus $\chi/\pi$.
  (a)~$F_{\nu r}^{(g)}/F_{\nu d}^{(g)}$;
  (b)~$F_{\nu r}^{(\mathrm{disk})}/F_{\nu d}^{(\mathrm{disk})}$.\label{fig:retdir}}
\end{figure}

Now it is interesting to compare the direct and retrograde scatterings. We show the ratios of the corresponding fluxes in Fig.~\ref{fig:retdir}. One can see that that the difference between the fluxes for the retrograde and direct scatterings can be about 20\%. This asymmetry remains valid for both the only gravitational scattering, shown in Fig.~\ref{7a}, and when the matter contribution is accounted for, which is depicted in Fig.~\ref{7b}.

\section{Discussion\label{sec:DISC}}

In the present work, we have studied spin oscillations in the neutrino
scattering off BHs. Both nonrotating and rotating BHs have been discussed. The neutrino spin evolution in curved
spacetime has been accounted for quasiclassically basing on the approach
developed in Refs.~\cite{Dvo06,Dvo13}. As an application of the obtained results, we have examined the neutrino
scattering off SMBH with a realistic accretion disk.

In Sec.~\ref{sec:GRAV}, we have derived the general expressions for the transition and survival probabilities for neutrino spin oscillations. Then, in Sec.~\ref{sec:SCHWARZ}, we have applied these results for the neutrino scattering off a nonrotating BH. The expression for the spin rotation angle, which is valid for the arbitrary neutrino energy and the impact parameter, has been presented in Eqs.~\eqref{eq:aobslim} and~\eqref{eq:eqtosolve}. Using this result, we have obtained that an ultrarelativistic neutrino preserves its helicity while scattering in the Schwarzschild metric. This feature is valid for any impact parameter $y\geq y_0$. This our finding is in agreement with the results of Ref.~\cite{DolDorLas06}. The transition probabilities $P_\mathrm{LR}$ of spin oscillations of massive neutrinos in their gravitational scattering turn out to be nonzero, and are shown in Fig.~\ref{fig:PLRSchwarz} for several neutrino energies. However, $P_\mathrm{LR}$ is rather small for reasonable Lorentz factors $E/m$ to be observed.

Then, in Sec.~\ref{sec:KERR}, we have considered spin oscillations of neutrinos scattered off a rotating BH. As in Sec.~\ref{sec:SCHWARZ}, we have derived the general formula for the spin rotation angle in Eqs.~\eqref{eq:alphaintKerr}-\eqref{eq:RKerr}. The obtained expressions reproduce the results of Sec.~\ref{sec:SCHWARZ} in the limit $z\to 0$. The transition probabilities of spin oscillations are nonzero even for ultrarelativistic neutrinos. In Fig.~\ref{fig:PLRKerr}, we have shown $P_\mathrm{LR}$ for the direct and retrograde scatterings of ultrarelativistic neutrinos for different angular momenta of BH.

Note that
the fact that the helicity of ultrarelativistic, or even massless, fermions
can be changed under the influence of a gravitational field was noticed
earlier in Refs.~\cite{Mer95,SinMobPap04}. We have found that spin oscillations of ultrarelativistic neutrinos happen only when particles scatter off a rotating BH. The polarization change of ultrarelativistic particles (photons) in their gravitational scattering off a rotating star was studied previously in Ref.~\cite{Gua02}.

Then, in Sec.~\ref{sec:MATT}, we have derived the effective Schr\"{o}dinger
equation for a neutrino scattering off BH surrounded by background
matter with a nonuniform density. In the case of only the gravitational
scattering, studied in Secs.~\ref{sec:SCHWARZ} and~\ref{sec:KERR}, it was possible to obtain the 
analytical transition and survival probabilities for some impact
parameters. If, besides gravity, a neutrino interacts with a background
matter, the probabilities can be derived only in the numerical solution
of Eq.~(\ref{eq:Schr}).

The effective Hamiltonians for neutrino spin oscillations for both the Schwarzschild and Kerr metrics have been obtained in Sec.~\ref{sec:MATT} in the approximation of a slowly rotating accretion disk. The effective Hamiltonian for ultrarelativistic neutrinos scattering off a rotating BH surrounded by an accretion disk has been also derived in Sec.~\ref{sec:MATT}.

In Sec.~\ref{sec:APPL}, we have considered the astrophysical applications
of our results. In particular, we have studied the effect of spin
oscillations on the neutrino scattering off SMBH surrounded by an
accretion disk. We have taken the parameters of the accretion disk,
such as the maximal number density and the profile of the mass distribution, close to the
values resulting from observations and hydrodynamics simulations.

First, we have studied the case of a nonrotating BH. Using the numerical solution of Eqs.~(\ref{eq:Schr}) and~(\ref{eq:OmegaxSchw}),
we have found the observed fluxes of outgoing neutrinos for only the gravitational scattering and when the neutrino interaction with the accretion disk is accounted for. The contribution of spin oscillations to the neutrino fluxes, shown in Figs.~\ref{fig:FgF0Schwarz} and~\ref{fig:Schwarzmatt}, appears to be negligible for reasonable neutrino energies and the current upper bound on the neutrino mass.

Greater spin effects have been revealed for neutrinos gravitationally scattered off a rotating BH surrounded by an accretion disk. In this situation, we have studied the ultrarelativistic neutrinos scattering off a maximally rotating SMBH to highlight the effect of spin oscillations. We have shown in Fig.~\ref{fig:FgF0Kerr} that the observed fluxes of gravitationally scattered neutrinos can be reduced by almost 10\%. The contribution of the neutrino interaction with matter changes the fluxes by about 5\%, see Fig.~\ref{fig:Kerrmatt}.

As one can see in Figs.~\ref{fig:FgF0Schwarz}-\ref{fig:Kerrmatt},
there is no deviation of the fluxes for the forward neutrino scattering
at $\chi=0$ if one compares them with the fluxes of scalar particles.
It means that neutrino spin oscillations do not affect the size of a BH shadow. The major effect of spin oscillations is for the backward
neutrino scattering at $\chi=\pi$. Thus the intensity of the glory
flux for ultrarelativistic neutrinos is almost 10\% less than for scalar particles in case a rotating BH.

The influence of the plasma interaction on the gravitational scattering
of photons was thoroughly studied previously~\cite{CunHer18}. The photons propagation
in plasma surrounding a nonrotating BH was examined in Ref.~\cite{PerTsu17}, where it was found that the form of the BH shadow is not changed. However its size
can be enlarged. The shape of the shadow of a rotating BH can be deformed~\cite{PerTsu17}.
Using Fig.~\ref{4b}, we conclude that there is an asymmetry
in the observed neutrino fluxes depending on the orientation of the
neutrino trajectory with respect to a slim accretion disk. This asymmetry
is maximal for the backward neutrino scattering. Unfortunately, this effect is rather small in the Schwarzschild metric. The asymmetry in the outgoing neutrino fluxes may well exist for the Kerr BH surrounded by an accretion disk. Basing on our results, we cannot quantitatively describe this effect since we rely on the equatorial neutrino motion only.

We have found that spin oscillations in the gravitational
scattering off a rotating BH are sizable for ultrarelativistic neutrinos. Thus our results are of interest for the neutrino astronomy~\cite{Sta09}, which is a rapidly developing
area of the cosmic rays physics. It is known that neutrinos with energies in the PeV range
were detected~\cite{Aar13}. Moreover, several sources of ultrahigh energy neutrinos can be identified with with some astronomical objects such as active galactic nuclei~\cite{Aar19}.
We have demonstrated in our work that, if the incoming flux of cosmic
neutrinos experience the gravitational lensing, in some cases, the observed flux
can be reduced by down to 10\%, compared to its initial value, because
of neutrino spin oscillations.

There is another possible application of the obtained results. The $r$-process nucleosynthesis in the vicinity of BH, surrounded by an accretion disk, was found in Ref.~\cite{WanJan12} to be affected by the neutrino radiation of the accretion disk. To influence the weak nuclear reactions, which the  nucleosynthesis is based on, emitted neutrinos should be active. In the present work, we predict the significant conversion of left active neutrinos to right sterile particles in the neutrino scattering off a rotating BH. Thus the nucleosynthesis near such BHs is further modified because of the neutrino gravitational interaction. Note that the effect of active to sterile neutrinos oscillations on the dynamics of the supernova explosion was studied in Ref.~\cite{HidFul07}.
 
\section*{Acknowledgments}
I am thankful to V.~I.~Dokuchaev and A.~V.~Yudin for useful comments.
This work is performed within the government assignment of IZMIRAN.

\appendix

\section{Particle motion in the Schwarzschild metric\label{sec:PARTM}}

In this Appendix, we briefly remind how to describe the motion of
a scalar particle interacting with a nonrotating BH, as well as how
to calculate the differential cross section of the gravitational scattering.
These problems were studied in details in Refs.~\cite[pp.~287-290]{LanLif71} and~\cite{DolDorLas06}.

The energy $E$ and the angular momentum $L$ are the conserved quantities
for a particle with the mass $m$ moving in the Schwarzschild metric
in Eq.~(\ref{eq:intschw}). The equation of motion and the trajectory
are defined by
\begin{align}\label{eq:eqmtr}
  \frac{\mathrm{d}r}{\mathrm{d}t}= & \pm\frac{m}{E}
  \left(
    1 - \frac{r_g}{r}
  \right)
  \left[
    \frac{E^{2}}{m^{2}}-
    \left(
      1 - \frac{r_g}{r}
    \right)
    \left(
      1+\frac{L^{2}}{m^{2}r^{2}}
    \right)
  \right]^{1/2},
  \notag
  \\
  \frac{\mathrm{d}\phi}{\mathrm{d}r}= & \pm\frac{L}{mr^{2}}
  \left[
    \frac{E^{2}}{m^{2}}-
    \left(
      1 - \frac{r_g}{r}
    \right)
    \left(
      1+\frac{L^{2}}{m^{2}r^{2}}
    \right)
  \right]^{-1/2},
\end{align}
for a particle moving in the equatorial plane. In Eq.~(\ref{eq:eqmtr}),
the minus signs stay for incoming particles and the plus ones for
outgoing particles.

The angle corresponding to the minimal distance between a particle and BH is
\begin{equation}\label{eq:minangle}
  \phi_{m}=y\int_{x_{m}}^{\infty}\frac{\mathrm{d}x}{\sqrt{xR_\mathrm{S}(x)}},
\end{equation}
where $y=b/r_{g}$, $b=L/E\sqrt{1-\gamma^{-2}}$ is the impact parameter, $R_\mathrm{S}(x)$ is given in Eq.~\eqref{eq:eqtosolve}, and $x_{m}$
is the maximal root of Eq.~\eqref{eq:eqtosolve}. The parameter $y>y_{0}$, where $y_0$ is given in Eq.~\eqref{eq:ycrit}.
Otherwise a particle falls to BH.

While computing the differential cross section, $\mathrm{d}\sigma/\mathrm{d}\varOmega$,
where $\mathrm{d}\varOmega=2\pi\sin\chi\mathrm{d}\chi$, we should
take into account that a particle, before being scattered off, can
make multiple revolutions around BH, both clockwisely and anticlockwisely.
One should account for this fact in the determination of the angle
$\chi$, fixing the position of a detector, which is in the range
$0<\chi<\pi$.

\begin{figure}
  \centering
  \includegraphics[scale=0.4]{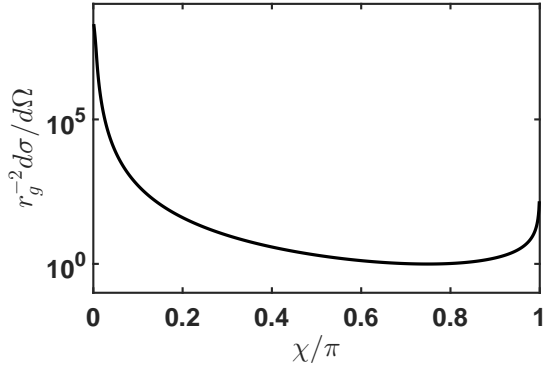}
  \caption{The differential cross section of the gravitational scattering of
  scalar particles off a nonrotating BH, normalized by $r_g^2$, versus $\chi/\pi$.
  The particle energy is $E=10m$.
  \label{fig:diffsc0}}
\end{figure}

In Fig.~\ref{fig:diffsc0}, we present the result of the numerical
computation of the cross section. While building this plot, we take
that $y_{0}<y<30y_{0}$ and account for up to two revolutions of a
particle around BH in both directions. This our result is used in
Sec.~\ref{sec:APPL} when we study the neutrino scattering off a
realistic BH surrounded by an accretion disk.

\section{Particle motion in the Kerr metric\label{sec:PARTMKERR}}

In this Appendix, we describe the motion of a scalar particle in the gravitational field of a rotating BH.

The trajectory of a particle in the Kerr metric, given in Eqs.~\eqref{eq:Kerrmetr} and~\eqref{eq:Kerrmetrparam}, is rather complicated in the general case. However, if we consider the particle motion in the equatorial plane, the trajectory and the equation of motion can be expressed in the relatively simple form (see, e.g., Ref.~\cite{Rez16}),
\begin{align}\label{eq:eqmtrKerr}
  \frac{\mathrm{d}t}{\mathrm{d}r} = &
  \pm \frac{\sqrt{r}}{r(r-r_{g})+a^{2}}
  \notag
  \\
  & \times
  \frac{[r^{3}+a^{2}(r_{g}+r)]E-a L r_{g}}
  {\sqrt{[r^{3}+a^{2}(r_{g}+r)]E^{2}-2aLEr_{g}-(r-r_g)L^{2}-m^{2}r[r(r-r_{g})+a^{2}]}},
  \notag
  \\
  \frac{\mathrm{d}\phi}{\mathrm{d}r} = &
  \pm \frac{\sqrt{r}}{r(r-r_{g})+a^{2}}
  \notag
  \\
  & \times
  \frac{(r-r_g)L+a E r_{g}}
  {\sqrt{[r^{3}+a^{2}(r_{g}+r)]E^{2}-2aLEr_{g}-(r-r_g)L^{2}-m^{2}r[r(r-r_{g})+a^{2}]}},
\end{align}
where one has two integrals of motion: the particle angular momentum $L$ and its energy $E$.

Analogously to Eq.~\eqref{eq:minangle}, we get the expression for $\phi_{m}$,
\begin{equation}
  \phi_{m} = \int_{x_{m}}^{\infty}
  \frac{\sqrt{x}\mathrm{d}x}{x(x-1)+z^{2}}
  \frac{(x-1)y\sqrt{1-\gamma^{-2}} + z}{\sqrt{(1-\gamma^{-2})R_\mathrm{K}(x)}},
\end{equation}
where $R_\mathrm{K}(x)$ is given in Eq.~\eqref{eq:RKerr}, and $x_m$ is the maximal root of Eq.~\eqref{eq:RKerr}.

\begin{figure}
  \centering
  \subfigure[]
  {\label{9a}
  \includegraphics[scale=.2]{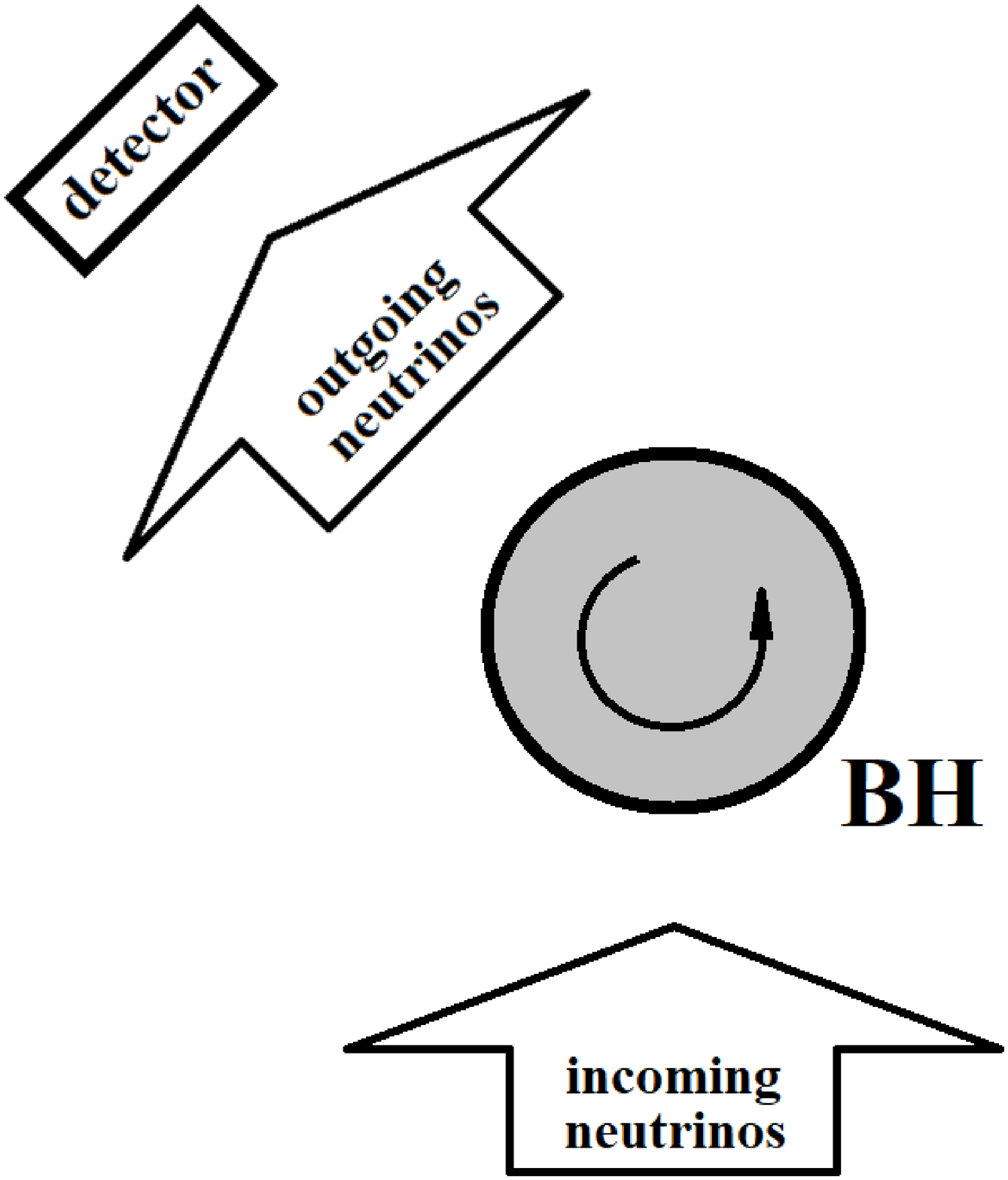}}
  \subfigure[]
  {\label{9b}
  \includegraphics[scale=.2]{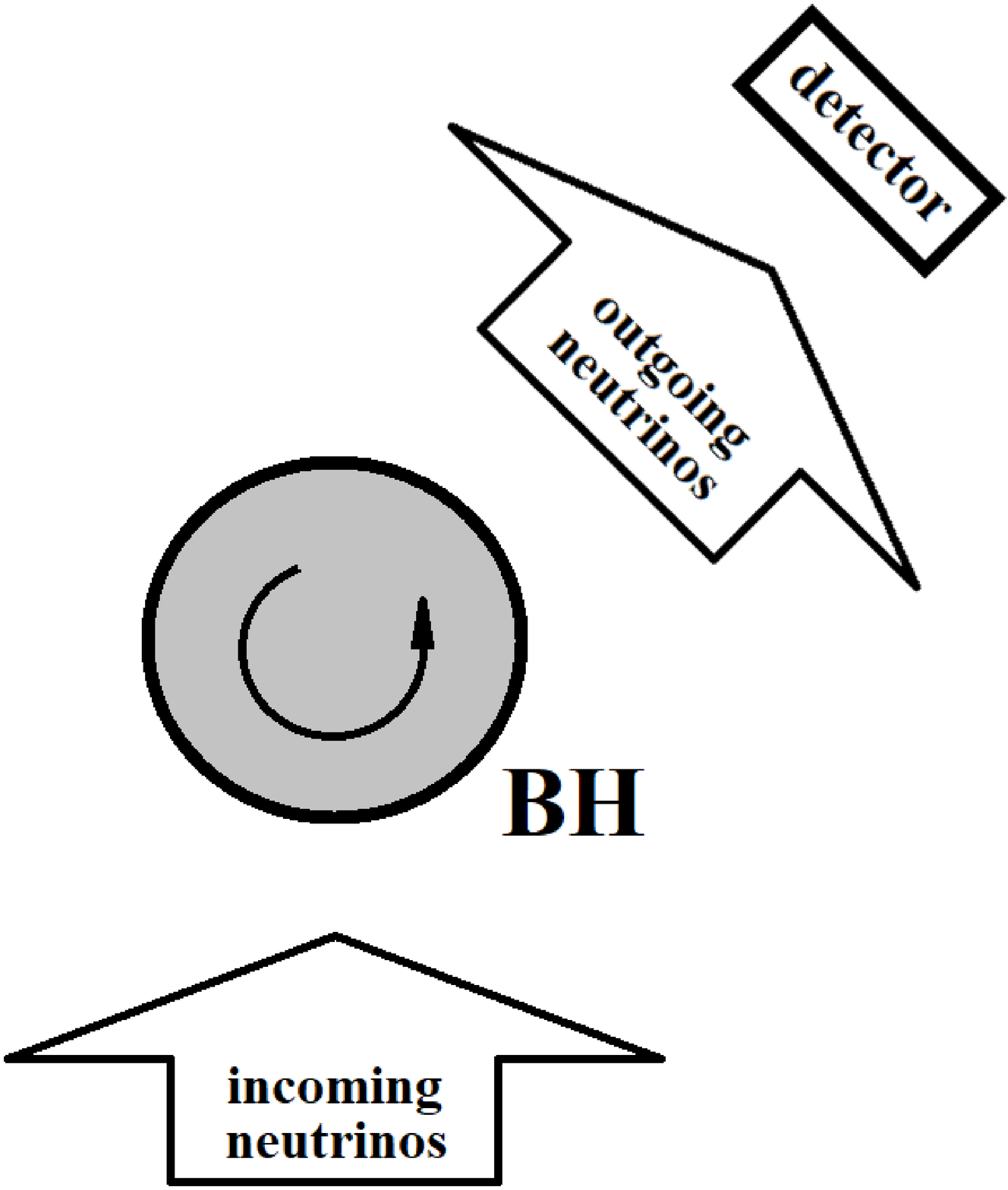}}
  \protect
  \caption{The illustration of the particle scattering off a rotating BH.
  (a)~Direct scattering; (b)~retrograde scattering.\label{fig:Kerrscatt}}
\end{figure}

There is an important difference between the particle scattering in the Schwarzschild and Kerr metrics. We can take that $L>0$ in the Schwarzschild metric. In the Kerr metric, the cases $L>0$ and $L<0$ are different. It can be illustrated in Fig.~\ref{fig:Kerrscatt}. One can call the situation, when the position of a detector is bent towards the BH rotation direction, as the direct scattering. It is shown in Fig.~\ref{9a}. The opposite situation, depicted in Fig.~\ref{9b} can be called the retrograde scattering. The corresponding quantities for the retrograde scattering can be obtained by replacing the signs in all terms with odd powers of $y$.

Of course, the total flux of, e.g., directly scattered particles contains contributions with $L<0$. Indeed, despite $0<\chi<\pi$ in Fig.~\ref{9a}, particles, which make one full revolution around BH, move oppositely the BH rotation, i.e. they have $L<0$. This fact should be accounted for while computing the total flux of scattered particles.

The calculation of the differential cross section of the gravitational scattering of scalar particles off a Kerr BH is analogous to the Schwarzschild metric studied in Appendix~\ref{sec:PARTM}. Thus we omit the details.


\begin{thebibliography}{50}

\bibitem{Aga18}
  N.~Agafonova \textit{et al.} (OPERA Collaboration),
  Final Results of the OPERA Experiment on $\nu_\tau$
  Appearance in the CNGS Neutrino Beam,
  Phys. Rev. Lett. \textbf{120}, 211801 (2018)
  [arXiv:1804.04912].

\bibitem{Abe18}
  K.~Abe \textit{et al.} (Super-Kamiokande Collaboration),
  Atmospheric neutrino oscillation analysis with external constraintsin Super-Kamiokande I-IV,
  Phys. Rev. D \textbf{97}, 072001 (2018)
  [arXiv:1710.09126].

\bibitem{Gir18}
  C.~Giganti, S.~Lavignac, and M.~Zito,
  Neutrino oscillations: The rise of the PMNS paradigm,
  Prog. Part. Nucl. Phys. \textbf{98}, 1--54 (2018)
  [arXiv:1710.00715].

\bibitem{FujShr80}
  K.~Fujikawa and R.~Shrock,
  Magnetic Moment of a Massive Neutrino and Neutrino Spin Rotation,
  Phys. Rev. Lett. \textbf{45}, 963--966 (1980).


\bibitem{GiuStu15}
  C.~Giunti, K.~A.~Kouzakov, Y.-F.~Li, A.~V.~Lokhov, A.~I.~Studenikin, and S.~Zhou,
  Electromagnetic neutrinos in laboratory experiments and astrophysics,
  Ann. Phys. (Amsterdam) \textbf{528}, 198--215 (2016)
  [arXiv:1506.05387].

\bibitem{Bil18}
  S. Bilenky,
  \textit{Introduction to the Physics of Massive and Mixed Neutrinos}, 2nd ed.
  (Springer, Cham, 2018).



\bibitem{Dvo06}
  M.~Dvornikov,
  Neutrino spin oscillations in gravitational fields,
  Int. J. Mod. Phys. D \textbf{15}, 1017\textendash 1034 (2006)
  [hep-ph/0601095].

\bibitem{Dvo13}
  M.~Dvornikov,
  Neutrino spin oscillations in matter under the influence of gravitational
  and electromagnetic fields,
  J. Cosmol. Astropart. Phys. 06 (2013) 015
  [arXiv:1306.2659].

\bibitem{Dvo19}
  M.~Dvornikov,
  Neutrino spin oscillations in external fields in curved spacetime,
  Phys. Rev. D \textbf{99}, 116021 (2019)
  [arXiv:1902.11285].


\bibitem{ObuSilTer17}
  Y.~N.~Obukhov, A.~J.~Silenko, and O.~V.~Teryaev,
  General treatment of quantum and classical spinning particles in external fields,
  Phys. Rev. D \textbf{96}, 105005 (2017)
  [arXiv:1708.05601].

\bibitem{SorZil07}
  F.~Sorge and S.~Zilio,
  Neutrino spin flip around a Schwarzschild black hole,
  Classical Quantum Gravity \textbf{24}, 2653\textendash 2664 (2007).

\bibitem{AlaNod15}
  S.~A.~Alavi and S.~Nodeh,
  Neutrino spin oscillations in gravitational fields in noncommutative spaces,
  Phys. Scripta \textbf{90}, 035301 (2015)
  [arXiv:1301.5977].


\bibitem{CroGiuMor04}
  R.~M.~Crocker, C.~Giunti, and D.~J.~Mortlock,
  Neutrino interferometry in curved spacetime,
  Phys. Rev. D \textbf{69}, 063008 (2004)
  [hep-ph/0308168].

\bibitem{AleClo18}
  J.~Alexandre and K.~Clough,
  Black  hole  interference  patterns  in  flavor  oscillations,
  Phys. Rev. D \textbf{98}, 043004 (2018)
  [arXiv:1805.01874].

\bibitem{Aki19}
  K.~Akiyama \textit{et al.} (Event Horizon Telescope Collaboration),
  First M87 event horizon telescope results.
  I. The shadow of the supermassive black hole,
  Astrophys. J. Lett. \textbf{875}, L1 (2019).

\bibitem{Abb19}
  B.~P.~Abbott \textit{et al.} (LIGO Scientific Collaboration  and Virgo Collaboration), 
  GWTC-1:  A gravitational-wave transientcatalog of compact binary mergers observed
  by LIGO and Virgo during the first and second observing runs,
  Phys. Rev. X \textbf{9}, 031040 (2019)
  [arXiv:1811.12907].

\bibitem{DokNazSmi19}
  V.~I.~Dokuchaev, N.~O.~Nazarova, and V.~P.~Smirnov,
  Event horizon silhouette:
  Implications to supermassive black holes in the galaxies M87 and Milky Way,
  Gen. Relativ. Gravit. \textbf{51}, 81 (2019)
  [arXiv:1903.09594].



\bibitem{WanJan12}
  S.~Wanajo and H.-Th.~Janka,
  The $r$-process in the neutrino-driven wind from a black-hole torus,
  Astrophys. J. \textbf{746}, 180 (2012)
  [arXiv:1106.6142].

\bibitem{Cor15}
  C.~Corian\`o, A.~Costantini, M.~Dell'Atti, and L.~Delle Rose,
  Neutrino and photon lensing by black holes:
  radiative lens equations and post-Newtonian contributions,
  J. High Energy Phys. 07 (2015) 160
  [arXiv:1504.01322].

\bibitem{StuSch19}
  Z.~Stuchl\'{i}k and J.~Schee,
  Shadow of the regular Bardeen black holes and comparison of the motion of photons
  and neutrinos,
  Eur. Phys. J. C \textbf{79}, 44 (2019).

\bibitem{CunHer18}
  P.~V.~P.~Cunha and C.~A.~R.~Herdeiro,
  Shadows and strong gravitational lensing: a brief review,
  Gen. Relativ. Gravit. \textbf{50}, 42 (2018)
  [arXiv:1801.00860].

\bibitem{LanLif71}
  L.~D.~Landau and E.~M.~Lifschitz,
  \textit{The Classical Theory of Fields}, 3rd ed.
  (Pergamon Press, Oxford, 1971).

\bibitem{DolDorLas06}
  S.~Dolan, C.~Doran, and A.~Lasenby,
  Fermion scattering by a Schwarzschild black hole,
  Phys. Rev. D \textbf{74}, 064005 (2006)
  [gr-qc/0605031].

\bibitem{Dvo20}
  M.~Dvornikov,
  Spin effects in neutrino gravitational scattering,
  Phys. Rev. D \textbf{101}, 056018 (2020)
  [arXiv:1911.08317].

\bibitem{Rez16}
  L.~Rezzolla,
  An introduction to astrophysical black holes and their dynamical production,
  in \textit{Astrophysical Black Holes},
  ed. by F.~Haardt, V.~Gorini, U.~Moschella, A.~Treves, and M.~Colpi
  (Springer, Cham, 2016), pp.~24--29.

\bibitem{MohPal04}
  R.~N.~Mohapatra and P.~B.~Pal,
  \textit{Massive Neutrinos in Physics and Astrophysics}, 3rd ed.
  (World Scientific, Singapore, 2004), p.~98.

\bibitem{DvoStu02}
  M.~Dvornikov and A.~Studenikin,
  Neutrino spin evolution in presence of general external fields,
  J. High Energy Phys. 09 (2002) 016
  [hep-ph/0202113].

\bibitem{Igu00}
  I.~V.~Igumenshchev, M.~A.~Abramowicz, and R.~Narayan,
  Numerical simulations of convective accretion flows in three dimensions,
  Astrophys. J. \textbf{537}, L27\textendash L30 (2000).

\bibitem{Jia19}
  J.~Jiang, A.~C.~Fabian, T.~Dauser, L.~Gallo, J.~A.~Garcia, E.~Kara, M.~L.~Parker, 
  J.~A.~Tomsick, D.~J.~Walton, and C.~S.~Reynolds,
  High Density Reflection Spectroscopy \textendash{}
  II. The density of the inner black hole accretion disc in AGN,
  Mon. Not. R. Astron. Soc. \textbf{489}, 3436\textendash 3455 (2019)
  [arXiv:1908.07272].

\bibitem{Ake19}
  M.~Aker \textit{et al.} (KATRIN Collaboration),
  Improved upper limit on the neutrino mass from a direct kinematic method by KATRIN,
  Phys. Rev. Lett. \textbf{123}, 221802 (2019)
  [arXiv:1909.06048].

\bibitem{Mer95}
  C.~Mergulh\~{a}o Jr.,
  Neutrino helicity flip in a curved space-time,
  Gen. Relativ. Gravit. \textbf{27}, 657\textendash 667 (1995).

\bibitem{SinMobPap04}
  D.~Singh, N.~Mobed, and G.~Papini,
  Helicity precession of spin-1/2 particles in weak inertial and gravitational fields,
  J. Phys. A: Math. Gen. \textbf{37}, 8329\textendash 8347 (2004)
  [hep-ph/0405296].

\bibitem{Gua02}
  E.~Guadagnini,
  Gravitational deflection of light and helicity asymmetry,
  Phys. Lett. B \textbf{548}, 19--23 (2002)
  [gr-qc/0207036].

\bibitem{PerTsu17}
  V.~Perlick and O.~Yu.~Tsupko,
  Light propagation in a plasma on Kerr spacetime:
  Separation of the Hamilton-Jacobi equation and calculation of the shadow,
  Phys. Rev. D \textbf{95}, 104003 (2017)
  [arXiv:1702.08768].

\bibitem{Sta09}
  T.~Stanev,
  \textit{High Energy Cosmic Rays}, 2nd ed.
  (Praxis Publishing, Chichester, 2009), pp.~298--313.


\bibitem{Aar13}
  M.~G.~Aartsen \textit{et al.} (IceCube Collaboration),
  First observation of PeV-energy neutrinos with IceCube,
  Phys. Rev. Lett. \textbf{111}, 021103 (2013)
  [arXiv:1304.5356].

\bibitem{Aar19}
  M.~G.~Aartsen \textit{et al.} (IceCube Collaboration),
  Time-integrated neutrino source searches with 10 years of IceCube data,
  Phys. Rev. Lett. \textbf{124}, 051103 (2020)
  [arXiv:1910.08488].

\bibitem{HidFul07}
  J.~Hidaka and G.~M.~Fuller,
  Sterile neutrino-enhanced supernova explosions,
  Phys. Rev. D \textbf{76}, 083516 (2007)
  [arXiv:0706.3886].

\end{thebibliography}
\end{document}